\documentclass[aps,prl,onecolumn,showpacs,preprintnumbers]{revtex4}
\usepackage{amsmath}
\usepackage{dcolumn}
\usepackage{bm}
\usepackage{subfigure}
\usepackage{amsfonts}
\usepackage{amssymb}
\usepackage{makeidx}
\usepackage{epsfig}
\usepackage{graphicx}

\begin{document}

\title{Hamiltonian approach to GR - Part 2: covariant theory of quantum
gravity}
\author{Claudio Cremaschini}
\affiliation{Institute of Physics and Research Center for Theoretical Physics and
Astrophysics, Faculty of Philosophy and Science, Silesian University in
Opava, Bezru\v{c}ovo n\'{a}m.13, CZ-74601 Opava, Czech Republic}
\author{Massimo Tessarotto}
\affiliation{Department of Mathematics and Geosciences, University of Trieste, Via
Valerio 12, 34127 Trieste, Italy}
\affiliation{Institute of Physics, Faculty of Philosophy and Science, Silesian University
in Opava, Bezru\v{c}ovo n\'{a}m.13, CZ-74601 Opava, Czech Republic}
\date{\today }

\begin{abstract}
A non-perturbative quantum field theory of General Relativity is presented
which leads to a new realization of the theory of Covariant Quantum-Gravity
(CQG-theory). The treatment is founded on the recently-identified
Hamiltonian structure associated with the classical space-time, i.e., the
corresponding manifestly-covariant Hamilton equations and the related
Hamilton-Jacobi theory. The quantum Hamiltonian operator and the CQG-wave
equation for the corresponding CQG-state and wave-function are realized in $%
4-$scalar form. The new quantum wave equation is shown to be equivalent to a
set of quantum hydrodynamic equations which warrant the consistency with the
classical GR Hamilton-Jacobi equation in the semiclassical limit. A
perturbative approximation scheme is developed, which permits the adoption
of the harmonic oscillator approximation for the treatment of the
Hamiltonian potential. As an application of the theory, the stationary
vacuum CQG-wave equation is studied, yielding a stationary equation for the
CQG-state in terms of the $4-$scalar invariant-energy eigenvalue associated
with the corresponding approximate quantum Hamiltonian operator. The
conditions for the existence of a discrete invariant-energy spectrum are
pointed out. This yields a possible estimate for the graviton mass together
with a new interpretation about the quantum origin of the cosmological
constant.
\end{abstract}

\maketitle

\section{1- Introduction}

This paper is part of a research project on the foundations of classical and
quantum gravity. Following the theoretical premises presented in Ref.\cite%
{noi3} (hereon referred to as Part 1), in this paper the axiomatic setting
of the theory of manifestly-covariant quantum gravity is established for the
Standard Formulation of General Relativity (briefly SF-GR), namely the
Einstein field equations and the corresponding treatment of the
gravitational field \cite{ein1,LL,gravi,wald}. The new quantum theory, based
on the introduction of massive gravitons and constructed in such a way to be
consistent with SF-GR in the semiclassical limit, will be referred to here
as theory of Covariant Quantum Gravity (CQG) or briefly \emph{CQG-theory}.

Distinctive features of CQG-theory presented here are that, just like the
CCG-theory (\textit{i.e.}, the theory of Covariant Classical Gravity)
developed in Part 1, it realizes a canonical quantization approach for SF-GR
which satisfies the principles of general covariance and manifest
covariance. This means that - in comparison with customary literature
canonical quantization approaches to SF-GR \cite{ADM,hh1} - the theory
proposed here preserves its form under arbitrary local point
transformations. As a consequence, in particular, it does not rely on the
adoption of space-time coordinates involving 3+1 or 2+2 foliation schemes
\cite{ADM,zzz2,alcu,Vaca2,Vaca5,Vaca6} (see also related discussion in Part
1).

In addition, first it is based on the adoption of $4-$tensor\ continuum
Lagrangian coordinates and canonical momentum operators and a
manifestly-covariant quantum wave equation referred to here as\ \emph{%
CQG-wave equation}. Second, the same quantum wave equation satisfies the
quantum unitarity principle and admits a closed set of equivalent quantum
hydrodynamic equations. Third, its formulation is of general validity,
\textit{i.e.}, it applies to arbitrary possible realizations of the
underlying classical space-time.

The goal of the present paper is also to display its non-perturbative
character which, nevertheless, allows for the development of perturbative
approximation schemes for the analytical evaluation of quantum solutions of
physical interest. The latter should include, in principle, also the
investigation of particular quantum solutions which can be considered
suitably "close", \textit{i.e}., localized, with respect to the background
classical space-time field tensor.

The theoretical framework is provided by Part 1 where\ a realization
inspired by the DeDonder-Weyl manifestly-covariant approach \cite%
{donder,weyl,sym9} was reached for the covariant Hamiltonian structure of
SF-GR $\left\{ x_{R},H_{R}\right\} $ based on CCG-theory. Its crucial
feature is that of relying upon the adoption of a new kind of variational
approach denoted as \emph{synchronous variational principle }earlier
developed in Refs.\cite{noi1,noi2}. According to the notations adopted in
Part 1, this involves the introduction of a variational set $\left\{
x_{R},H_{R}\right\} $, referred to as the\ classical GR-Hamiltonian
structure, formed respectively by a suitable reduced-dimensional classical
canonical\ state $x_{R}$\ and a corresponding manifestly-covariant
Hamiltonian density $H_{R}$. More precisely, the variational canonical state
$x_{R}$ is identified with the set $x_{R}=\left\{ g,\pi \right\} $, with%
\textbf{\ }$g\equiv \left\{ g_{\mu \nu }\right\} $\textbf{\ }and $\pi \equiv
\left\{ \pi _{\mu \nu }\right\} $\ respectively representing the continuum
Lagrangian coordinate and conjugate canonical momentum, both realized by
means of second-order $4-$tensors. Instead, consistent with Ref.\cite{noi2}
and in formal analogy with the customary symbolic representation holding in
relativistic particle dynamics, the Hamiltonian density $H_{R}$ is taken of
the form%
\begin{equation}
H_{R}\ \equiv T_{R}+V,  \label{hamiltonian density}
\end{equation}%
with $T_{R}$ and $V$ denoting suitable $4-$scalar effective "kinetic" and
"potential" densities. As shown in Part 1, these fields, together with the
state $x_{R}$, are then prescribed in such a way to provide an appropriate\
Hamiltonian variational formulation of the Einstein field equations, \textit{%
i.e}., to yield a corresponding equivalent set of continuum\ Hamilton
equations.

A characteristic\ physical requirement of the resulting Hamiltonian theory
is that it should satisfy the \emph{general covariance principle} (GCP) with
respect\ to local point transformations \cite{noi4} as well as its more
restrictive manifest covariant form, namely the \emph{principle of manifest
covariance} (PMC).\ Accordingly it should always be possible to represent in
$4-$tensor form all the relevant field variables and operators, including
the variational functional, the corresponding Lagrangian and Hamiltonian
densities and operators, the canonical variables as well their variations
and the Euler-Lagrange equations.

The prescription of the covariant Hamiltonian structure $\left\{
x_{R},H_{R}\right\} $ consistent with these properties, as well as the
construction of the corresponding manifestly-covariant Hamilton-Jacobi
theory for SF-GR developed in Part 1, are \emph{mandatory physical
prerequisites} for the establishment of the CQG-theory.

Meeting these physical requirements appears "\textit{a priori}" a difficult
task despite the huge number of contributions to be found in the past
literature and dealing with Quantum Gravity. In fact,\ one has to notice
that a common difficulty met by many of previous non-perturbative
Hamiltonian approaches to GR is that they are based on the adoption of
variational Lagrangian densities, Lagrangian coordinates and/or momenta
which have a non-manifestly covariant character, \textit{i.e}., they are not
$4-$scalars or $4-$tensors. Incidentally, a strategy of this type is
intrinsic in the construction of the original Einstein's variational
formulation for his namesake field equations\textbf{\ }which is based on the
Einstein-Hilbert asynchronous variational principle (see related discussion
in Ref.\cite{noi1}). A choice of\ analogous type, for example, is typical of
the Dirac's Hamiltonian approach to GR, which is based on the so-called
Dirac's constrained dynamics \cite{dirac4,dirac3,cast1,cast2,cast3}. Its key
principle, in fact, is that of singling out the \textquotedblleft
time\textquotedblright\ component of the $4-$position in terms of which the
generalized velocity is identified with $g_{\mu \nu ,0}$. This lead him to
identify the canonical momentum in terms of the manifestly non-tensorial
quantity $\pi _{Dirac}^{\mu \nu }=\frac{\partial \mathcal{L}_{EH}}{\partial
g_{\mu \nu ,0}}$, where $\mathcal{L}_{EH}$ is the Einstein-Hilbert
variational Lagrangian density.\ Such a choice corresponds to select a
particular subset of GR-frames.

In the cases indicated above the possibility is prevented of establishing at
the classical level a Hamiltonian theory of SF-GR in which the
Euler-Lagrange equations (in particular the Hamilton equations) are
manifestly covariant. On the other hand, the conjecture that a
manifestly-covariant Hamiltonian formulation must be possible for continuum
systems is also suggested by the analogous theory holding for discrete
classical particle systems. Indeed, its validity is fundamentally implied
by\ the state-of the-art theory of classical $N-$body systems subject to
non-local electromagnetic (EM) interactions. The issue is exemplified by the
Hamiltonian structure of the EM radiation-reaction problem in the case of
classical extended particles as well as $N-$body EM interactions among
particles of this type \cite{EPJ1,EPJ2,EPJ3,EPJ4,EPJ5}. Nevertheless, it
must be mentioned that in the case of continuum fields, the appropriate
formalism is actually well-established, being provided by the Weyl-DeDonder
Lagrangian and Hamiltonian treatments \cite{donder,weyl,sym9}. The need to
adopt an analogous approach also in the context of classical GR, and in
particular for the Einstein equation itself or its possible modifications,
has been recognized before \cite%
{esposito1995,sym6,sym7,rovelli2002,rovelli2003}. The fulfillment of the
physical prerequisites indicated above in the context of a classical
treatment of SF-GR and the definition of the related conceptual framework
for GR has been provided recently by Part 1 and Refs.\cite{noi1,noi2}.

The viewpoint adopted in this paper for the development of the new approach
to the covariant quantum gravity has analogies with the \textit{%
first-quantization} approach developed in Ref.\cite{EPJP-2015}. This
pertains to the relativistic quantum theory of an extended charged particle
subject to EM self-interaction, the so-called EM radiation reaction, and
immersed in a flat Minkowski space-time. In fact, as shown in the same
reference, the appropriate relativistic quantum wave equation advancing the
quantum state of the same particle was achieved: first, by the preliminary
construction of the manifestly-covariant classical Hamilton and
Hamilton-Jacobi equation for the classical dynamical system \cite{EPJ7};
second, by the classical treatment of external and self interactions acting
on the quantum system, both expressed in terms of deterministic force
fields, so to leave the development of quantum theory only for the particle
state dynamics.

As shown here an analogous procedure can be adopted in the case of covariant
quantum gravity in order to achieve the proper form of the relativistic
quantum wave equation corresponding to SF-GR. More precisely, consistent
with the setting required by the adoption of synchronous variational
principle, a so-called \emph{background space-time picture} is adopted, so
that the underlying classical space-time geometry $(Q^{4},\widehat{g}(r))$\
is considered as prescribed, in the sense that when it is parametrized in
terms of arbitrary curvilinear coordinates $r\equiv \left\{ r^{\mu }\right\}
$\ its metric tensor $\widehat{g}_{\mu \nu }(r)$\ is regarded as a \emph{%
prescribed classical field},\textbf{\ }eventually to be identified with the
metric tensor of the physical space-time. Such a choice is of crucial
importance since it permits to recover "\textit{habitual physical notions
such as causality, time, scattering states, and black holes}" \cite{hh1}.

For this purpose, the construction of the covariant quantum wave equation
for the gravitational field reached in this paper is based on the classical
GR-Hamilton-Jacobi equation reported in Part 1, in which the\emph{\
prescribed field }$\widehat{g}_{\mu \nu }(r)$ is assumed to realize a
particular smooth solution of the Einstein field equations and to determine
in this way the \emph{geometric structure} associated with the background
space-time $\left( \mathbf{Q}^{4},\widehat{g}\left( r\right) \right) $.
Hence, $\widehat{g}_{\mu \nu }(r)$ establishes the tensor transformation
laws by raising and lowering indexes of all tensor quantities, together with
the prescribed values of the invariant space-time volume element as well as
of the standard connections (Christoffel symbols), covariant derivatives and
Ricci tensor. It must be stressed that, despite the adoption of the
prescribed field $\widehat{g}_{\mu \nu }(r)$ in the construction of the
quantum theory of gravity reported here, the whole treatment remains "%
\textit{a priori}" \textit{exact}, \textit{i.e}., non-asymptotic in
character, and at the same time \textit{background-independent}, since the
theory does not rely on a particular realization of $\widehat{g}_{\mu \nu
}(r)$, which can be any solution of the Einstein field equations.

In the framework of quantum theory the\ prescription of the\ background
geometric structure has also formal conceptual analogies with the so-called
induced gravity (or emergent gravity), namely the conjecture that the
geometrical properties of space-time reveal themselves as a suitable mean
field description of microscopic stochastic or quantum degrees of freedom
underlying the classical solution. In the present approach this is achieved
by introducing in the Lagrangian and Hamiltonian operators themselves the
notion of prescribed metric tensor $\widehat{g}_{\mu \nu }\left( r\right) $\
which is held constant under the action of all the quantum operators and has
therefore to be distinguished from the quantum field $g_{\mu \nu }$. As a
result, the classical variational field $g_{\mu \nu }$\ is now interpreted
as a quantum observable. The ensemble spanned by all possible values of $%
g_{\mu \nu }$\ determines the configuration space $U_{g}$\ with respect to
which the quantum-gravity state has to be prescribed, so that $U_{g}$\ can
be identified with the real vector space $U_{g}\equiv
\mathbb{R}
^{16}$ (or $U_{g}\equiv
\mathbb{R}
^{10}$ if the quantum field $g_{\mu \nu }$ is assumed symmetric).

Taking into account these considerations the work-plan of the paper is as
follows. In Section 2, the principles of the axiomatic formulation of
manifestly-covariant quantum gravity corresponding to the
reduced-dimensional GR-Hamiltonian structure earlier reported in Ref.\cite%
{noi3} are discussed, with the aim of addressing in particular the following
Axioms of CQG:\newline
\emph{CQG -Axiom 1:} prescription of the quantum gravity state (CQG-state) $%
\psi $, to be assumed a $4-$scalar complex function of the form $\psi =\psi
(g,\widehat{g},s)$, with $g,\widehat{g}\in U_{g}\subseteq
\mathbb{R}
^{16}$ and $s\in I\equiv
\mathbb{R}
$, with $\psi $ spanning a Hilbert space $\Gamma _{\psi }$, \textit{i.e}., a
finite dimensional linear vector space endowed with a scalar product to be
properly prescribed. Here $s$\ denotes the proper time along an arbitrary
background geodetics, \textit{i.e}., prescribed requiring that the line
element $ds$\ satisfies the differential identity $ds^{2}=\widehat{g}_{\mu
\nu }(r(s))dr^{\mu }dr^{\nu }$, with $dr^{\mu }$ being the tangent
displacement performed along the same geodetics.\newline
\emph{CQG -Axiom 2: }prescription of the expectation values of the quantum
observables and of the related quantum probability density function
(CQG-PDF).\newline
\emph{CQG -Axiom 3: }formulation of the quantum correspondence principle for
the GR-Hamiltonian structure. This includes the prescription of the form of
the quantum Hamiltonian operator generating the proper-time evolution of the
CQG-state.

In Section 3 the problem is posed of the prescription of a quantum wave
equation (CQG-wave equation), namely the formulation of the quantum wave
equation advancing in proper time the same CQG-state (\emph{CQG -Axiom 4}).
As for the classical theory developed in Part 1, the covariant quantization
of the gravitational field reached here is realized in a $4-$dimensional
space-time. Then it is shown that, upon introducing a Madelung
representation for the quantum wave function and invoking validity of the
quantum unitarity principle, the CQG-wave equation is equivalent to a couple
of quantum hydrodynamic equations identified respectively with the
continuity and quantum Hamilton-Jacobi equations (\emph{CQG -Axiom 5}). In
particular, given validity of the semiclassical limit, the CQG-wave equation
is proved to recover the classical Hamilton-Jacobi equation reported in Part
1, thus warranting the conceptual consistency between the two descriptions.
In Section 4 the development of a perturbative approach to CQG-theory
starting from the exact quantum representation is presented, a feature that
allows for the implementation of the harmonic oscillator approximation for
the analytical treatment of the quantum Hamiltonian potential.

A number of selected applications of CQG-theory are considered. In
particular, in Section 5 the investigation of the stationary CQG-wave
equation holding in the case of vacuum and subject to the assumption of
having a non-vanishing cosmological constant is treated. Then, in Section 6
the proof of the existence of a discrete spectrum for the energy eigenvalues
associated with the same vacuum CQG-wave equation is obtained. In Section 7,
based on the identification of\ the minimum energy eigenvalue of the
discrete spectrum for the vacuum CQG-wave equation, the quantum prescription
of the rest-mass $m_{o}$ as well as of the corresponding characteristic
scale length\ $L(m_{o})$ entering the CQG-theory are discussed. This
provides an estimate for the ground-state graviton mass of the vacuum
solution, which is proved to be strictly positive under the same physical
prescriptions which warrant the existence of a discrete energy spectrum. At
the same time it yields a new interpretation of the quantum origin of the
cosmological constant, which is shown to be related to the Compton
wavelength of the ground-state oscillation mode of the quantum of the
gravitational field. In Section 8, main differences and comparisons with the
two main existing categories of literature approaches to Quantum Gravity,
which are usually referred to as canonical and covariant quantization
approaches respectively, are pointed. Finally, in Section 9 the main
conclusions and a summary of the investigation are presented.

\section{2 - Axiomatic foundations of CQG - Axioms 1-3}

In this section we start addressing the axiomatic formulation of CQG which
is consistent with the physical prerequisites indicated above. In particular
here the axioms are provided which permit to prescribe the key physical
properties of the\emph{\ quantum GR-Hamiltonian system} associated\ with the
GR-Hamiltonian structure $\left\{ x_{R},H_{R}\right\} $, or equivalently its
dimensional-normalized representation $\left\{ \overline{x}_{R},\overline{H}%
_{R}\right\} $, both specified in Part 1. These include:\newline
- \emph{The functional setting of the quantum gravity state (CQG-state) }$%
\psi $.\newline
- \emph{The definition of quantum expectation values and observables.}%
\newline
- \emph{The correspondence principle between the classical and quantum
GR-Hamiltonian systems, to be established in terms of a mapping between the
classical canonical state and the related classical GR-Hamiltonian density}
\emph{with the corresponding quantum observables.}\newline

\subsection{2A - The Hilbert space of the CQG\textbf{-state} $\protect\psi $}

The first Axiom concerns the prescription of the \emph{CQG-state} and its
corresponding functional space to be identified with a Hilbert space.

\emph{CQG -Axiom 1 -} \emph{The Hilbert space of the CQG-state }$\psi $

\emph{The CQG-state is identified with a single }$4-$\emph{scalar and
complex function\ }$\psi (s)$\emph{\ (CQG wave-function) of the form}%
\begin{equation}
\psi (s)\equiv \psi (g,\widehat{g}(r),r(s),s).  \label{Axiom-1-1}
\end{equation}%
\emph{This can be associated with a corresponding spin-}$2$\emph{\ quantum
particle having a strictly-positive invariant rest mass }$m_{o}$\emph{. In
fact, in the context of a first-quantization approach developed here, }$\psi
(s)$\emph{\ can always be identified with the tensor product of the form }$%
\psi (s)=\widehat{g}_{\mu \nu }\psi ^{\mu \nu }$\emph{. Regarding the
functional setting of }$\psi (s)$\emph{\ here it is assumed that:}

A$_{1}$)\emph{\ }$\psi (s)$\emph{\ is taken to depend smoothly on the tensor
field }$g\equiv \left\{ g_{\mu \nu }\right\} $ \emph{spanning the
configurations space} $U_{g}$ \emph{and in addition to admit a Lagrangian
path (LP) parametrization in terms of the\ geodetics }$r(s)\equiv \left\{
r^{\mu }(s)\right\} $\emph{\ associated locally with the prescribed
classical tensor field }$\widehat{g}(r)\equiv \left\{ \widehat{g}_{\mu \nu
}(r)\right\} $\emph{. It means that it may be smoothly-dependent on the
prescribed field in terms of the parametrization }$\widehat{g}(r(s))\equiv
\left\{ \widehat{g}_{\mu \nu }(r(s))\right\} ,$ \emph{on the }$s-$\emph{%
parametrized geodetics\ }$r(s)\equiv \left\{ r^{\mu }(s)\right\} $\emph{\
and on the classical proper time }$s$\emph{\ associated with the same
geodetics.}

B$_{1}$)\emph{\ The functions }$\psi $\emph{\ defined by Eq.(\ref{Axiom-1-1}%
) span a Hilbert space }$\Gamma _{\psi },$ \emph{\textit{i.e}., a
finite-dimensional linear vector space endowed with the scalar product}%
\begin{equation}
\left\langle \psi _{a}|\psi _{b}\right\rangle \equiv
\int\limits_{U_{g}}d(g)\psi _{a}^{\ast }(g,\widehat{g}(r),r(s),s)\psi _{b}(g,%
\widehat{g}(r),r(s),s),  \label{Axiom 1-2}
\end{equation}%
\emph{with }$d(g)\equiv \prod\limits_{\mu ,\nu =1,4}$ $dg_{\mu \nu }$\emph{\
denoting the canonical measure on }$U_{g}$\emph{\ and }$\psi _{a,b}(s)\equiv
\psi _{a,b}(g,\widehat{g}(r),r(s),s)$\emph{\ being arbitrary elements of the
Hilbert space }$\Gamma _{\psi }$\emph{, where as usual }$\psi _{a}^{\ast }$%
\emph{\ denotes the complex conjugate of }$\psi _{a}$\emph{.}

C$_{1}$) \emph{The real function }$\rho (s)\equiv \rho (g,\widehat{g}%
(r),r(s),s)$ \emph{prescribed as }%
\begin{equation}
\rho (s)\equiv \left\vert \psi (s)\right\vert ^{2}  \label{Axiom-1-3}
\end{equation}%
\emph{identifies on the configuration space }$U_{g}$ \emph{the quantum
probability density function (CQG-PDF) associated with the CQG-state. Here
by assumption }$\rho (s)$\emph{\ is the probability density of }$g\equiv
\left\{ g_{\mu \nu }\right\} $\emph{\ in the volume element }$d(g)$\emph{\
belonging to the configuration space }$U_{g}.$\emph{\ Thus, if }$L_{g}$
\emph{is an arbitrary subset of }$U_{g}$\emph{, its probability is defined as%
}%
\begin{equation}
P(L_{g})=\int\limits_{U_{g}}d(g)\rho (s)\Theta (L_{g}),
\end{equation}%
\emph{with }$\Theta (L_{g})$\emph{\ denoting the characteristic function of }%
$L_{g},$\emph{\ namely such that }$\Theta (L_{g})=1,0$\emph{\ if
respectively }$g\equiv \left\{ g_{\mu \nu }\right\} \in U_{g}$\emph{\
belongs or not to }$L_{g}.$\emph{\ In addition, by assumption the
normalization }%
\begin{equation}
P(U_{g})\equiv \left\langle \psi |\psi \right\rangle \equiv
\int\limits_{U_{g}}d(g)\rho (s)=1  \label{normalization}
\end{equation}%
\emph{is assumed to hold identically for arbitrary }$(\widehat{g}(r),r(s),s)$%
\emph{.}

D$_{1}$) \emph{The real function }$S^{(q)}(s)\equiv S^{(q)}(g,\widehat{g}%
(r),r(s),s)$\emph{\ defined as}%
\begin{equation}
S^{(q)}(s)\equiv \arcsin \text{h }\left\{ \frac{\psi (s)-\psi ^{\ast }(s)}{%
2\rho (s)}\right\}  \label{Axiom 1-4}
\end{equation}%
\emph{identifies on the configuration space }$U_{g}$ \emph{the quantum
phase-function} \emph{associated with the same CQG-state }$\psi (s)$\emph{.}

In Eqs.(\ref{Axiom-1-1}), (\ref{Axiom 1-2}) and (\ref{Axiom-1-3}) given
above the notations are as follows. First, $g\equiv \left\{ g_{\mu \nu
}\right\} $ is the continuum Lagrangian coordinate spanning the
configuration space $U_{g}\subseteq
\mathbb{R}
^{16}$. Second, $\widehat{g}(r)\equiv \left\{ \widehat{g}_{\mu \nu
}(r)\right\} $ is the classical deterministic $4-$tensor which for an
arbitrary coordinate parametrization $r\equiv \left\{ r^{\mu }\right\} $
identifies the prescribed metric tensor of the background space-time $\left(
\mathbf{Q}^{4},\widehat{g}(r)\right) $, which lowers (and raises) the tensor
indices of all tensor fields. Third, $r(s)$ $\equiv \left\{ r^{\mu }\right\}
$ is the $4-$position belonging to the local geodesics associated with the
prescribed metric tensor $\widehat{g}(r)$ such that for an arbitrary $r\in
\left( \mathbf{Q}^{4},\widehat{g}(r)\right) $ locally occurs that $r\equiv
r(s)$. Fourth, $s$ is the proper time on the same local geodesics $\left\{
r(s)\right\} $ which spans the time axis $I\equiv
\mathbb{R}
$.

\subsection{2B - Expectation values and observables}

The second Axiom deals with the prescription of the expectation values of
tensor functions and CQG-observables, namely configuration-space $4-$tensor
functions or $4-$tensor operators whose expectation values are expressed in
terms of real tensor functions.

\emph{CQG -Axiom 2 -} \emph{Prescription of CQG-expectation values and
CQG-observables}

\emph{Given an arbitrary tensor function or local tensor operator} $%
X(s)\equiv X(g,\widehat{g}(r),r(s),s)$ \emph{which acts on an arbitrary
wave-function }$\psi (s)\equiv \psi (g,\widehat{g}(r),r(s),s)$ \emph{of the
Hilbert space }$\Gamma _{\psi }$\emph{,} \emph{the weighted integral}%
\begin{equation}
\left\langle \psi |X\psi \right\rangle \equiv \int\limits_{U_{g}}d(g)\psi
^{\ast }(s)X(s)\psi (s),  \label{Axiom-2-1}
\end{equation}%
\emph{which is assumed to exist, is denoted as\ CQG-expectation value of }$X$%
\emph{.\ Then by construction }$\left\langle \psi |X\psi \right\rangle $%
\emph{\ is a }$4-$\emph{tensor field, generally of the form}%
\begin{equation}
\left\langle \psi |X\psi \right\rangle =G_{X}(\widehat{g}(r),r(s),s).
\label{Axiom.2.2}
\end{equation}%
\emph{In the particular case in which }$\left\langle \psi |X\psi
\right\rangle $\emph{\ is real, namely}%
\begin{equation}
\left\langle \psi |X\psi \right\rangle =\left\langle X^{\ast }\psi |\psi
\right\rangle \equiv \int\limits_{U_{g}}d(g)\psi (s)X^{\ast }(s)\psi ^{\ast
}(s),
\end{equation}%
\emph{with }$X^{\ast }(s)$\emph{\ denoting the complex conjugate of }$X(s),$
\emph{then }$X$\emph{\ identifies a CQG-observable}.

The trivial example of observable is provided by the identification $X\equiv
1$. The normalization condition (\ref{normalization}) then necessarily
yields in such a case $G_{X}=1$. Other examples of CQG-observables include:

A) The $4-$tensor function $X\equiv g_{\mu \nu }$. Then, for all $r\equiv
r(s)\in \left( \mathbf{Q}^{4},\widehat{g}(r)\right) $ the integral%
\begin{equation}
\left\langle \psi |g_{\mu \nu }\psi \right\rangle
=\int\limits_{U_{g}}d(g)g_{\mu \nu }\rho (g,\widehat{g}(r),r(s),s)=%
\widetilde{g}_{\mu \nu }(\widehat{g}(r),r(s),s)  \label{Axiom-2-3}
\end{equation}%
identifies the CQG-expectation value of $g_{\mu \nu }$ at $r\equiv r(s)$.
Here $\widetilde{g}_{\mu \nu }(s)\equiv \widetilde{g}_{\mu \nu }(\widehat{g}%
(r),r(s),s)$ is by construction a real tensor field to be considered
generally different from the prescribed metric tensor $\widehat{g}(r)(\equiv
\widehat{g}(s))\equiv \left\{ \widehat{g}_{\mu \nu }(r(s))\right\} $, while $%
r\equiv r(s)\equiv \left\{ r^{\mu }(s)\right\} $ is again the $4-$position
of the background space-time $\left( \mathbf{Q}^{4},\widehat{g}(r)\right) $.
Thus, letting $\delta g_{\mu \nu }(r)=\widehat{g}_{\mu \nu }(r)-\widetilde{g}%
_{\mu \nu }(r)$ it follows that the CQG-expectation value of the tensor
function $X\equiv g_{\mu \nu }+\delta g_{\mu \nu }(r)$ is just%
\begin{equation}
\left\langle \psi |\left( g_{\mu \nu }+\delta g_{\mu \nu }(r)\right) \psi
\right\rangle =\int\limits_{U_{g}}d(g)\left( g_{\mu \nu }+\delta g_{\mu \nu
}(r)\right) \rho (g,\widehat{g}(r),r(s),s)=\widehat{g}_{\mu \nu }(r),
\label{Axiom-2-3-bis}
\end{equation}%
\textit{i.e}., it coincides identically with the deterministic classical
metric tensor which at the $4-$position $r\equiv r(s)$ is associated with $%
\left( \mathbf{Q}^{4},\widehat{g}(r)\right) $. In the light of the classical
theory developed in Part 1, the quantum expectation values provided by Eqs.(%
\ref{Axiom-2-3}) and (\ref{Axiom-2-3-bis}), \textit{i.e}., respectively $%
\widetilde{g}_{\mu \nu }(\widehat{g}(r),r(s),s)$\ and $\widehat{g}_{\mu \nu
}(r)$\ should be suitably related. This point will be discussed elsewhere.

B) The CQG-expectation value of the \emph{momentum CQG-operator }$\pi
^{(q)\mu \nu }\equiv -i\hslash \frac{\partial }{\partial g_{\mu \nu }}$ is
prescribed in such a way that the integral
\begin{equation}
\left\langle \psi |\pi ^{(q)\mu \nu }\psi \right\rangle
=\int\limits_{U_{g}}d(g)\psi ^{\ast }(g,\widehat{g}(r),r(s),s)\left(
-i\hslash \frac{\partial }{\partial g_{\mu \nu }}\right) \psi ^{\ast }(g,%
\widehat{g}(r),r(s),s)\equiv \widetilde{\pi }^{(q)\mu \nu }(\widehat{g}%
(r),r(s),s)
\end{equation}%
always exists, with $\widetilde{\pi }^{(q)\mu \nu }$ being a real tensor
field. Thus the CQG-operator $\pi ^{(q)\mu \nu }$ is necessarily a
CQG-observable. Then, introducing the CQG-operator $T\equiv \pi ^{(q)\mu \nu
}\pi _{\mu \nu }^{(q)}$, its CQG-expectation value is just
\begin{equation}
\left\langle \psi |T\psi \right\rangle =\int\limits_{U_{g}}d(g)\psi ^{\ast
}(g,\widehat{g}(r),r(s),s)T\psi ^{\ast }(g,\widehat{g}(r),r(s),s)\equiv
\widetilde{T}(\widehat{g}(r),r(s),s),  \label{t-oper}
\end{equation}%
which is assumed to exist, with $\widetilde{T}$ manifestly identifying a
real scalar field. Therefore the CQG-operator $T$ is necessarily a
CQG-observable too.

\subsection{2C - Prescription of the CQG-correspondence principle}

The classical dimensionally-normalized Hamiltonian structure $\left\{
\overline{x}_{R},\overline{H}_{R}\right\} $ determined in Part 1 is defined
in terms of the canonical state $\overline{x}_{R}\equiv \left\{ \overline{g}%
_{\mu \nu },\overline{\pi }_{\mu \nu }\right\} $ and the Hamiltonian $%
\overline{H}_{R}$. More precisely, $\overline{g}_{\mu \nu }\equiv g_{\mu \nu
}$ and $\overline{\pi }_{\mu \nu }=\frac{\alpha L}{k}\pi _{\mu \nu }$ is the
normalized conjugate momentum, where $\kappa =\frac{c^{3}}{16\pi G}$, $L$ is
a $4-$scalar scale length and $\alpha $ is a suitable dimensional $4-$%
scalar, both to be defined below. Instead, $\overline{H}_{R}$ is defined as
the real $4-$scalar field%
\begin{equation}
\overline{H}_{R}(\overline{x}_{R},\widehat{g},r,s)=\overline{T}_{R}\left(
\overline{g},\widehat{g},r,s\right) +\overline{V}\left( \overline{g},%
\widehat{g},r,s\right) ,  \label{HAmiltonian-N}
\end{equation}%
with $\overline{T}_{R}\left( \overline{g},\widehat{g},r,s\right) \equiv
\frac{1}{f(h)}\frac{\overline{\pi }^{\mu \nu }\overline{\pi }_{\mu \nu }}{%
2\alpha L}$ and $\overline{V}\left( \overline{g},\widehat{g},r,s\right)
\equiv \sigma \overline{V}_{o}\left( \overline{g},\widehat{g},r,s\right)
+\sigma \overline{V}_{F}\left( \overline{g},\widehat{g},r,s\right) $ being
the normalized effective kinetic and potential densities. Here $\overline{V}%
_{o}\equiv h\alpha L\left[ g^{\mu \nu }\widehat{R}_{\mu \nu }-2\Lambda %
\right] $ and $V_{F}\equiv \frac{h\alpha L}{2k}L_{F}$ represent respectively
the vacuum and external field contributions (see definitions in Part 1),
with $\widehat{R}_{\mu \nu }$ being the background Ricci tensor and $\Lambda
$ being the cosmological constant. Finally, $f(h)$\ and $\sigma =\pm 1$\ are
suitable\emph{\ multiplicative gauges,}\textbf{\ }\textit{i.e}., real $4-$%
scalar fields which remain in principle still arbitrary at the classical
level, where $h=\left( 2-\frac{1}{4}g^{\alpha \beta }(r)g_{\alpha \beta
}(r)\right) $ and $f(h)$\ satisfies by construction the constraint $f(%
\widehat{g}(r))=1$.

Given these premises, we can now introduce the core canonical quantization
rules in the context of CQG-theory. These are based on a suitable
correspondence principle between the classical and the relevant quantum
functions and operators. This is given as follows.

\emph{CQG -Axiom 3- CQG-correspondence principle for the GR-Hamiltonian
structure.}

\emph{Given the classical GR-Hamiltonian structure }$\left\{ \overline{x}%
_{R},\overline{H}_{R}\right\} $,\emph{\ the CQG-correspondence principle is
realized by the map}%
\begin{equation}
\left\{
\begin{array}{c}
\overline{g}_{\mu \nu }\equiv g_{\mu \nu }\rightarrow g_{\mu \nu
}^{(q)}\equiv g_{\mu \nu }, \\
\overline{\pi }_{\mu \nu }\rightarrow \pi _{\mu \nu }^{(q)}\equiv -i\hslash
\frac{\partial }{\partial g^{\mu \nu }}, \\
\overline{H}_{R}\rightarrow \overline{H}_{R}^{(q)}=\frac{1}{f(h)}\overline{T}%
_{R}^{(q)}(\overline{\pi })+\overline{V},%
\end{array}%
\right.  \label{map-3}
\end{equation}%
\emph{where} $\overline{H}_{R}^{(q)}$ \emph{is the CQG-Hamiltonian
operator,\ }$x^{(q)}\equiv \left\{ g_{\mu \nu }^{(q)},\pi _{\mu \nu
}^{(q)}\right\} $\emph{\ is the quantum canonical state and }$\pi _{\mu \nu
}^{(q)}$\emph{\ is the quantum momentum operator prescribed so that the
commutator }$\left[ g_{\mu \nu }^{(q)},\pi ^{(q)\alpha \beta }\right]
=i\hslash \delta _{\mu }^{\alpha }\delta _{\nu }^{\beta }$ \emph{exactly. In
addition, }$\overline{T}_{R}^{(q)}(\overline{\pi })$\emph{\ is the kinetic
density quantum operator}%
\begin{equation}
\overline{T}_{R}^{(q)}(\overline{\pi })=\frac{\pi ^{(q)\mu \nu }\pi _{\mu
\nu }^{(q)}}{2\alpha L}\equiv \frac{T}{2\alpha L},
\end{equation}%
\emph{where }$T\equiv \pi ^{(q)\mu \nu }\pi _{\mu \nu }^{(q)}$ \emph{is}
\emph{the }$4-$\emph{scalar operator introduced above (see Eq.(\ref{t-oper}%
)).}

Hence, Eqs.(\ref{map-3}) mutually map in each other respectively the
classical canonical state $\overline{x}_{R}$ and the GR-Hamiltonian density $%
\overline{H}_{R}$ onto the corresponding quantum variables/operators $%
x^{(q)} $ and $\overline{H}_{R}^{(q)}$. Given the prescriptions (\ref{map-3}%
), key consequence (of CQG-Axiom 3) is therefore the validity of the
canonical commutation rule at the basis of the canonical quantization
formalism of CQG theory, namely
\begin{equation}
\left[ \pi _{\mu \nu }^{(q)},g_{\alpha \beta }^{(q)}\right] =-i\hslash
\delta _{\mu \alpha }\delta _{\nu \beta }.  \label{canonical quantization}
\end{equation}%
It is worth pointing out that the same correspondence principle (\ref{map-3}%
) also prescribes the gauge properties of the quantum Hamiltonian operator $%
\overline{H}_{R}^{(q)}$\ and canonical momentum $\pi _{\mu \nu }^{(q)}$\
provided by gauge transformations of the corresponding classical fields $%
\overline{H}_{R}$\ and $\overline{\pi }_{\mu \nu }$ given in Part 1.\ In
particular, this means that $\overline{H}_{R}^{(q)}$\ and the canonical
momentum $\pi _{\mu \nu }^{(q)}$\ are endowed respectively with the gauge
transformations%
\begin{equation}
\left\{
\begin{array}{c}
\pi _{\mu \nu }^{(q)}\rightarrow \pi _{\mu \nu }^{\prime (q)}=f(h)\pi _{\mu
\nu }^{(q)}, \\
\overline{T}_{R}^{(q)}(\overline{\pi })\rightarrow \frac{1}{f(h)}\overline{T}%
_{R}^{(q)}(\overline{\pi }),%
\end{array}%
\right.  \label{quantum-gauge-1}
\end{equation}%
where $f(h)$\ denotes in principle an arbitrary, non-vanishing and
smoothly-differentiable $4-$scalar function, whose precise value is
determined below by requiring validity of quantum hydrodynamic equations in
conservative form.

\section{3 - Axiomatic foundations of CQG - Axioms 4 and 5}

In this Section\ additional constitutive aspects of CQG are formulated which
concern the prescription of:\newline
- \emph{The generic form of the CQG-wave equation advancing in proper time }$%
\psi $\emph{\ itself }(\emph{CQG-Axiom 4})\emph{.}\newline
- \emph{The realization of the corresponding quantum hydrodynamic equations
in conservative form} (\emph{CQG-Axiom 5}).

\subsection{3A - The quantum-gravity wave equation}

We first introduce the quantum wave equation which in the framework of CQG
realizes an evolution equation for the quantum state $\psi (s)$. This is
provided by the following axiom.

\emph{CQG -Axiom 4 - Prescription of the CQG-wave equation for }$\psi $

\emph{The evolution equation advancing in time the CQG-state }$\psi (s)$%
\emph{\ is assumed to be provided by the\ CQG-quantum wave equation (CQG-QWE)%
}%
\begin{equation}
i\hslash \frac{\partial }{\partial s}\psi (s)+\left[ \psi (s),\overline{H}%
_{R}^{(q)}\right] =0,  \label{QG-WAVW EQUATION}
\end{equation}%
\emph{where }$\left[ A,B\right] \equiv AB-BA$\emph{\ denotes the quantum
commutator, \textit{i.e}.,}%
\begin{equation}
\left[ \psi (s),\overline{H}_{R}^{(q)}\right] \equiv -\overline{H}%
_{R}^{(q)}\psi (s).
\end{equation}

The CQG-wave equation (\ref{QG-WAVW EQUATION})\textbf{\ }uniquely prescribes
the evolution of the quantum state $\psi (s)$ along the geodetics of the
prescribed metric tensor $\widehat{g}_{\mu \nu }(r)$ which is associated
with the background curved space-time $\left( \mathbf{Q}^{4},\widehat{g}%
_{\mu \nu }(r)\right) $. Equation (\ref{QG-WAVW EQUATION}) is a first order
partial differential equation, to be supplemented by suitable initial
conditions, namely prescribing for all $r(s_{o})=r_{o}\in \left( \mathbf{Q}%
^{4},\widehat{g}(r)\right) $ the condition $\psi (s_{o})=\psi _{o}(g,%
\widehat{g}(r_{o}),r_{o})$, as well as boundary conditions at infinity on
the improper boundary of configuration space $U_{g}$, \textit{i.e}., letting
$\lim_{g\rightarrow \infty }\psi (g,\widehat{g}(r),r(s),s)=0$.

Nevertheless, $\psi (s)$\ and in particular $\psi (s_{o})$\ are manifestly
non-unique. This is due to gauge property indicated above (see Eq.(\ref%
{quantum-gauge-1})) which characterizes the Hamiltonian and canonical
momentum operators $\overline{H}_{R}^{(q)}$ and $\pi _{\mu \nu }^{(q)}$.
However, in this regard, a potential consistency issue arises for Axiom 4.
More precisely, this refers to the compatibility of Eq.(\ref{QG-WAVW
EQUATION}) with the normalization condition (\ref{normalization}) earlier
set at the basis of Axiom 1. In fact it is not obvious whether Eq.(\ref%
{normalization}) may actually hold identically for all $s\in I\equiv
\mathbb{R}
$ for arbitrary choices of the Hamiltonian operator $\overline{H}_{R}^{(q)}$
and in particular arbitrary choices of the undetermined function $f(h)$ (see
the rhs of the third equation in Eqs.(\ref{map-3})). In fact, Eq.(\ref%
{normalization}) demands in such a case also the validity of the additional
requirement%
\begin{equation}
\frac{\partial }{\partial s}\int\limits_{U_{g}}d(g)\rho (s)\equiv
\int\limits_{U_{g}}d(g)\frac{\partial }{\partial s}\rho (s)\equiv 0
\label{normalization-2}
\end{equation}%
to hold identically for all $s\in I$. The issue will be addressed in Section
3B.

\subsection{3B - The quantum hydrodynamic equations}

Let us now investigate whether and under which conditions the\ CQG-wave
equation introduced in Axiom 4 (see Eq.(\ref{QG-WAVW EQUATION})) may be
equivalent to a prescribed set of quantum hydrodynamic equations (QHE)
written in conservative form, \textit{i.e}., in such a way to conserve
quantum probability. In fact,\ in analogy with the Schroedinger equation and
the generalized Klein-Gordon equation reported in Ref.\cite{EPJP-2015} for
the radiation-reaction problem, the QHE should be realized respectively by:
a) a continuity equation for the quantum PDF $\rho (s)$; b) a quantum
Hamilton-Jacobi equation for the quantum phase-function $%
S^{(q)}(s)=S^{(q)}(g,\widehat{g}(r),r(s),s)$. \emph{We remark that the
derivation of QHE is required since it provides a theoretical framework for
the physical prescription of the gauge indeterminacy on }$f(h)$ \emph{%
characterizing the CQG-wave equation and the logical consistency of the
CQG-theory with the classical Hamilton-Jacobi equation determined in Part 1.}

We notice preliminarily that the CQG-state defined by the complex function $%
\psi (s)$ (see Eq.(\ref{Axiom-1-1})) can always be cast in the form of an
exponential representation of the type\ realized by the Madelung
representation as%
\begin{equation}
\psi (s)=\sqrt{\rho (s)}\exp \left\{ \frac{i}{\hslash }S^{(q)}(s)\right\} ,
\label{Axiom-5-1}
\end{equation}%
being $\rho (s)$\ and $S^{(q)}(s)$\ the real $4-$scalar field functions
prescribed respectively by Eqs.(\ref{Axiom-1-3}) and (\ref{Axiom 1-4}). The
following additional Axiom is then introduced.

\emph{CQG -Axiom 5 -}\ \emph{Quantum hydrodynamics equations}

\emph{Given validity of the Madelung representation (\ref{Axiom-5-1}), in
terms of the Hamiltonian operator }$\overline{H}_{R}^{(q)}$\emph{, then\
provided the constraint condition}\textbf{\ }%
\begin{equation}
f(h)=1  \label{CONSTRAINT CONDITION}
\end{equation}%
\emph{is fulfilled in order to satisfy the quantum unitarity principle,%
\textbf{\ }the CQG-wave equation (\ref{QG-WAVW EQUATION}) is equivalent to
the following set of real PDEs }%
\begin{eqnarray}
\frac{\partial \rho (s)}{\partial s}+\frac{\partial }{\partial g_{\mu \nu }}%
\left( \rho (s)V_{\mu \nu }(s)\right) &=&0,  \label{QHE-fin-1} \\
\frac{\partial S^{(q)}(s)}{\partial s}+\overline{H}_{c}^{(q)} &=&0,
\label{QHE-fin-2}
\end{eqnarray}%
\emph{referred to here as\ CQG-quantum continuity equation and CQG-quantum
Hamilton-Jacobi equation advancing in proper-time respectively }$\rho (s)$\
\emph{and }$S^{(q)}(s).$ \emph{Here the notation is as follows. The quantum
hydrodynamics fields }$\rho (s)\equiv \rho (g,\widehat{g},s)$\emph{\ and }$%
S^{(q)}(s)\equiv S^{(q)}(g,\widehat{g},s)$\emph{\ are\ assumed to depend
smoothly on the tensor field }$g\equiv \left\{ g_{\mu \nu }\right\} $\emph{\
spanning the configurations space }$U_{g}$\emph{\ and in addition to admit a
LP-parametrization in terms of the\ geodetics }$r(s)\equiv \left\{ r^{\mu
}(s)\right\} $\emph{\ associated locally with the prescribed field }$%
\widehat{g}(r)\equiv \left\{ \widehat{g}_{\mu \nu }(r)\right\} $\emph{. The }%
$4-$\emph{tensor }$V_{\mu \nu }(s)$\emph{\ is prescribed as }$V_{\mu \nu
}(s)=\frac{1}{\alpha L}\frac{\partial S^{(q)}}{\partial g^{\mu \nu }}$.
\emph{Then, }$\overline{H}_{c}^{(q)}$\emph{\ identifies up to an arbitrary
multiplicative gauge transformation (see Eq.(\ref{quantum-gauge-1})) the
effective quantum Hamiltonian density}%
\begin{equation}
\overline{H}_{c}^{(q)}=\frac{1}{2\alpha L}\frac{1}{f(h)}\frac{\partial
S^{(q)}}{\partial g^{\mu \nu }}\frac{\partial S^{(q)}}{\partial g_{\mu \nu }}%
+V_{QM}+\overline{V},  \label{prescription-2}
\end{equation}%
\emph{with }$V\equiv \overline{V}(g,\widehat{g}(r),r,s)$\emph{\ being the
effective potential density and\ }$V_{QM}$\emph{\ a potential density
denoted as Bohm-like effective quantum potential which is prescribed as}%
\begin{equation}
V_{QM}(g,\widehat{g}(r),r,s)=\frac{\hslash ^{2}}{8\alpha L}\frac{\partial
\ln \rho }{\partial g^{\mu \nu }}\frac{\partial \ln \rho }{\partial g_{\mu
\nu }}-\frac{\hslash ^{2}}{4\alpha L}\frac{\partial ^{2}\rho }{\rho \partial
g_{\mu \nu }\partial g^{\mu \nu }}.  \label{prescription-3}
\end{equation}%
\emph{Then one can show that validity of Eqs.(\ref{QHE-fin-1}) and (\ref%
{QHE-fin-2}) necessarily requires to uniquely fix the arbitrary
multiplicative gauge function }$f(h)$\emph{\ in Eq.(\ref{map-3}) so that
identically in the prescription of the function }$\overline{H}_{c}^{(q)}$
\emph{given above} \emph{Eq.(\ref{CONSTRAINT CONDITION}) must be fulfilled.}

The proof of the statement follows from elementary algebra. One notices in
fact that, upon substituting Eq.(\ref{Axiom-5-1}) in Eq.(\ref{QG-WAVW
EQUATION}), explicit evaluation delivers respectively for arbitrary $s\in
I\equiv
\mathbb{R}
$:%
\begin{eqnarray}
\frac{\partial \rho (s)}{\partial s}+\frac{1}{f(h)}\frac{\partial }{\partial
g_{\mu \nu }}\left( \rho (s)V_{\mu \nu }(s)\right) &=&0,  \label{QHE-A1} \\
\frac{\partial S^{(q)}(s)}{\partial s}+\frac{1}{f(h)}\frac{1}{2\alpha L}%
\frac{\partial S^{(q)}(s)}{\partial g^{\mu \nu }}\frac{\partial S^{(q)}(s)}{%
\partial g_{\mu \nu }}+V_{QM}(s)+\overline{V}_{R}(s) &=&0,  \label{QHE-A2}
\end{eqnarray}%
where the first one coincides with Eq.(\ref{QHE-fin-1}) if Eq.(\ref%
{CONSTRAINT CONDITION}) is satisfied. Hence this implies necessarily that
also in equation (\ref{prescription-3}) which defines $\overline{H}%
_{c}^{(q)} $,\ $f(h)$\ must be determined in the same way. Incidentally one
notices also that the prescription for $f(h)$ given above is also consistent
with the normalization condition (\ref{normalization}) holding for the
CQG-PDF and in particular with Eq.(\ref{normalization-2}) too. Indeed,
integration of the continuity equation (\ref{QHE-fin-1}) manifestly recovers
identically Eq.(\ref{normalization-2}). Hence the constraint condition Eq.(%
\ref{CONSTRAINT CONDITION}) is actually required to warrant the \emph{%
quantum unitarity principle, }namely the conservation of quantum
probability,\ \textit{i.e}.,\ the validity of Eq.(\ref{normalization}) for
all $s\in I$. In addition, this uniquely determines also the Hamiltonian
structure holding at the classical level, \textit{i.e}., the precise form of
the variational Hamiltonian density $H_{R}$.

Important theoretical outcomes follow from the CQG-quantum Hamilton-Jacobi
equation determined here. The first one is that the same equation
generalizes the classical GR-Hamilton-Jacobi equation earlier determined in
Part 1 (see Eq.(32) in Part 1), so that - in analogy to the same equation
and Ref. \cite{FoP2} - it must imply the validity of corresponding Hamilton
equations to be expressed in terms of the effective quantum Hamiltonian
density $\overline{H}_{c}^{(q)}$ (see Eq.(\ref{prescription-2})).\textbf{\ }%
It follows that, due to the presence of the Bohm-like effective quantum
potential $V_{QM}(g,\widehat{g}(r),r,s),$ the latter now generally must
depend explicitly on the proper time $s$ (see also related discussion in
Part 1, Subsection 2D).\textbf{\ }Detailed implications, involving the
construction of time-dependent solutions of the non-stationary-dependent
CQG-wave equation (\ref{QG-WAVW EQUATION}), will be discussed elsewhere.

The second outcome concerns the validity of the so-called semiclassical
limit of CQG, to be prescribed letting $\hslash \rightarrow 0$. By requiring
that in the same limit both $\alpha $ and $L(m_{o})$ reduce to their
classical definition and that the real limit function $\lim_{\hslash
\rightarrow 0}\frac{S^{(q)}(s)}{\hslash }=\frac{S(s)}{\alpha }$ exists for
arbitrary $s\in I\equiv
\mathbb{R}
$, with $S(s)$\ identifying the classical reduced Hamilton principal
function\ (see Part 1), then one can shown that the quantum Hamilton-Jacobi
equation (\ref{QHE-fin-2}) reduces to the analogous classical
Hamilton-Jacobi equation (see Part 1), while the limit $\lim_{\hslash
\rightarrow 0}\frac{V_{QM}(s)}{\hslash }=0$ holds identically. In fact,
considering without loss of generality the case of vacuum, the semiclassical
limit of Eq.(\ref{QHE-A2}) delivers%
\begin{equation}
\frac{1}{\alpha }\frac{\partial S(s)}{\partial s}+\frac{1}{2\alpha ^{2}L}%
\frac{\partial S(s)}{\partial g^{\mu \nu }}\frac{\partial S(s)}{\partial
g_{\mu \nu }}+\lim_{\hslash \rightarrow 0}\frac{V_{QM}(s)}{\hslash }=0,
\label{evaluation H-J}
\end{equation}%
where by construction the last term on the lhs vanishes identically. As a
consequence the effective quantum Hamiltonian density\emph{\ }$\overline{H}%
_{c}^{(q)}$ necessarily must reduce to the limit function%
\begin{equation}
\overline{H}_{o}=\frac{1}{2\alpha L}\frac{\partial S(s)}{\partial g^{\mu \nu
}}\frac{\partial S\left( s\right) }{\partial g_{\mu \nu }}.
\end{equation}%
This coincides in form with the classical normalized Hamiltonian density
given above by Eq.(\ref{HAmiltonian-N}), while Eq.(\ref{evaluation H-J})
reduces to the classical GR-Hamilton-Jacobi equation. Hence, this proves
that the derivation of the quantum hydrodynamic equations is a fundamental
theoretical result to be established for the validity of CQG-theory. Indeed,
the CQG-quantum continuity equation prescribes the expression of the gauge
function $f(h)$, while the CQG-quantum Hamilton-Jacobi equation establishes
the connection of the CQG-wave equation with the classical Hamilton-Jacobi
theory determined in Part 1. This issue represents a necessary conceptual
consistency aspect of quantum theory to be ascertained to hold between
classical and quantum descriptions of gravitational field dynamics.

Finally, one notices that the effective potential $V_{QM}(s)$\ introduced
here (see Eq.(\ref{prescription-3})) is analogous to the well-known Bohm
potential met in non-relativistic quantum mechanics (see for example Refs.%
\cite{FoP1,FoP2}), its physical origin being similar and arising due to the
non-uniformity of the quantum PDF. In the present case the non-uniformity
occurs because generally it must be $\frac{\partial }{\partial g_{\mu \nu }}%
\rho (s)\neq 0$, with consequent non-vanishing contributions arising in Eqs.(%
\ref{QHE-fin-1})-(\ref{QHE-fin-2}).

\section{4 - Perturbative approximation scheme}

In this section a theoretical feature related to the CQG-wave equation
determined above is analyzed. The issue here is whether - based on suitable
asymptotic orderings - a perturbative scheme can be developed both for the
classical GR-Hamiltonian theory and the corresponding CQG-theory indicated
above, in order to allow the adoption of a harmonic oscillator approximation
for the analytical treatment of the Hamiltonian potential. To prove how this
goal can be reached we start considering the decomposition
\begin{equation}
g_{\mu \nu }=\widehat{g}_{\mu \nu }(r)+\delta g_{\mu \nu },
\label{DECOMPOSITION FOR g}
\end{equation}%
with $\delta g^{\mu \nu },$ referred to here as the \emph{displacement }$4-$%
\emph{tensor field}, to be assumed suitably small. Concerning the notation,
we remark that hereon the symbol\ $\delta g^{\mu \nu }$ is used to indicate
displacement of the $4-$tensor field $g_{\mu \nu }$ and must be
distinguished from the similar notation adopted in Part 1 which refers
instead to the synchronous variations of tensorial fields. Then\emph{\ }it
is obvious that $\delta g_{\mu \nu }$ identifies, both in the context of
classical and quantum theories, an equivalent possible realization of the
Lagrangian coordinates which is alternative to $g_{\mu \nu }$. To make
further progress, however, we need also an approximation scheme. For this
purpose $g_{\mu \nu }$ is required to belong to a suitable infinitesimal
neighborhood of $\widehat{g}(r)\equiv \left\{ \widehat{g}_{\mu \nu
}(r)\right\} $, \textit{i.e}., the subset of $U_{g}$ denoted as
\begin{equation}
U_{g}(\widehat{g}(r),\varepsilon )=\left\{ g_{\mu \nu }\equiv \widehat{g}%
_{\mu \nu }(r)+\delta g_{\mu \nu },\delta g_{\mu \nu }\lesssim O(\varepsilon
),g_{\mu \nu }\in U_{g}\right\} ,  \label{neighborhood}
\end{equation}%
such that for all displacements $\delta g_{\mu \nu }$ the asymptotic ordering%
\begin{equation}
\delta g_{\mu \nu }\lesssim O(\varepsilon )  \label{ORDERING-1}
\end{equation}%
holds. Here $\varepsilon $ is a suitable infinitesimal real parameter, while
by construction in such a set all components of $\delta g_{\mu \nu }$ are of
order $O(\varepsilon )$ or higher. Let us consider the implications of Eqs.(%
\ref{DECOMPOSITION FOR g}) and (\ref{ORDERING-1}) in the two cases and
applying them - for definiteness - in validity of the prescription $%
f(h)\equiv 1$.

First, in the case of the classical Hamiltonian theory one notices that in $%
U_{g}(\widehat{g}(r),\varepsilon )$ the normalized GR-Hamilton equations
(see Part 1) can be equivalently represented as%
\begin{equation}
\left\{
\begin{array}{c}
\frac{D\delta \overline{g}_{\mu \nu }}{Ds}=\frac{\overline{\pi }_{\mu \nu }}{%
\alpha L}, \\
\frac{D\overline{\pi }_{\mu \nu }}{Ds}=-\frac{\partial \overline{V}%
^{(a)}\left( g,\widehat{g}\right) }{\partial \overline{\delta g}^{\mu \nu }},%
\end{array}%
\right.  \label{ASYMPTOTIC-REN-HAM}
\end{equation}%
with $\overline{V}^{(a)}\left( g,\widehat{g}\right) $ being the potential to
be conveniently approximated. When the identity $\overline{V}\left( g,%
\widehat{g}\right) \equiv \overline{V}_{o}\left( g,\widehat{g}\right) $
holds, elementary algebra yields up to an additive constant gauge the
asymptotic approximation
\begin{equation}
\overline{V}\left( g,\widehat{g}\right) \cong \frac{\sigma \alpha L}{4}%
\left\{ -\left[ \delta g_{\alpha \beta }\delta g^{\alpha \beta }\widehat{g}%
^{\mu \nu }+2\widehat{g}_{\alpha \beta }\delta g^{\alpha \beta }\delta
g^{\mu \nu }\right] \widehat{R}_{\mu \nu }+2\Lambda \delta g_{\mu \nu
}\delta g^{\mu \nu }\right\} \equiv \overline{V}_{R}^{(a)}\left( g,\widehat{g%
}\right) .
\end{equation}%
Then, thanks to the vacuum solution with non-vanishing cosmological constant
discussed in Part 1, for which $\widehat{R}_{\mu \nu }=\Lambda \widehat{g}%
_{\mu \nu }$, the previous equation delivers:%
\begin{equation}
\overline{V}_{R}^{(a)}\left( g,\widehat{g}\right) \equiv -\frac{\sigma
\alpha L\Lambda }{2}\left\{ \delta g_{\alpha \beta }\delta g^{\alpha \beta }+%
\widehat{g}_{\alpha \beta }\delta g^{\alpha \beta }\delta g^{\mu \nu }%
\widehat{g}_{\mu \nu }\right\} ,  \label{VACUUM POTENTIAL EXPANDED}
\end{equation}%
$\overline{V}_{R}^{(a)}\left( g,\widehat{g}\right) $ representing the vacuum
normalized effective potential density in the same infinitesimal
neighborhood indicated above. An analogous approximation holding for the
non-vacuum case can readily be obtained. It must be remarked that in all
cases the conceptual consistency underlying the harmonic expansion of the
Hamiltonian potential is granted by the structural stability analysis of the
classical Hamiltonian theory performed in Part 1 and the related
determination of the conditions for the occurrence of stable classical
solutions.

In the context of CQG-theory the transformation of the Lagrangian
coordinates $g_{\mu \nu }\rightarrow \delta g_{\mu \nu }$ is manifestly
obtained replacing the correspondence principle realized by means of Eqs.(%
\ref{map-3}) with the equivalent map
\begin{equation}
\left\{
\begin{array}{c}
\delta g_{\mu \nu }\rightarrow \delta g_{\mu \nu }^{(q)}\equiv \delta g_{\mu
\nu }, \\
\overline{\pi }_{\mu \nu }\rightarrow \pi _{\mu \nu }^{(q)}=-i\hslash \frac{%
\partial }{\partial \delta g^{\mu \nu }}\equiv \delta \pi _{\mu \nu }, \\
\overline{H}_{R}\rightarrow \overline{H}_{R}^{(q)}=\frac{\pi ^{(q)\mu \nu
}\pi _{\mu \nu }^{(q)}}{2\alpha L}+\overline{V},%
\end{array}%
\right.  \label{pass}
\end{equation}%
Notice that in principle\ the asymptotic ordering (\ref{ORDERING-1}) may
affect, in some sense, also the behavior of the quantum PDF\textbf{\ }$\rho
(s)=\rho (\widehat{g}(r)+\delta g,\widehat{g}(r),r(s),s)$. Indeed in the
limit $\delta g_{\mu \nu }\rightarrow 0,$ and consequently also if $\delta
g_{\mu \nu }$ is considered as an infinitesimal of $O(\varepsilon )$, it
means that the probability density $\rho (s)$\ should be suitably localized
in the set $U_{g}(\widehat{g}(r),\varepsilon )$\ indicated above.
Nevertheless, due to the arbitrariness of the solutions which the CQG-wave
equation may have, the possibility of prescribing "\textit{a priori}" its
precise asymptotic behavior seems unlikely.

\section{5 - \emph{Application \#1} - Construction of the stationary vacuum
CQG-wave equation}

The CQG-wave equation (\ref{QG-WAVW EQUATION}) admits generally
non-stationary solutions $\psi (s)\equiv \psi (g,\widehat{g},r,s)$, \textit{%
i.e}., in which the proper-time dependence cannot be simply factored out.
This happens because the CQG wave equation admits generally solutions which
are far from the classical one, \textit{i.e}., the prescribed background
solution $\widehat{g}_{\mu \nu }(r)$ (which is stationary by assumption, see
also Part 1).\ For sure they are important for applications/developments of
the theory, although the proper treatment of such issues must be forcibly
left to a separate investigation.

However, an equally important issue is to investigate the possible existence
of a discrete spectrum associated with the stationary CQG-wave equation
following from Eq.(\ref{QG-WAVW EQUATION}).\textbf{\ }Let us consider for
this purpose the case of vacuum in which by construction $\overline{V}=%
\overline{V}_{o}$, requiring that the cosmological constant $\Lambda $ and
the still arbitrary multiplicative gauge constant $\sigma $ are such that $%
-\sigma \Lambda >0$, and hence either $\left( \Lambda >0,\sigma =-1\right) $
or $\left( \Lambda <0,\sigma =1\right) $. For definiteness, let us assume
that the CQG-Hamiltonian operator $\overline{H}_{R}^{(q)}$\emph{\ }defined
by the quantum correspondence principle (see Eqs.(\ref{map-3})) does not
depend on $s$, at least in an asymptotic sense, so that the CQG-wave
equation admits exact (or asymptotic) separable particular solutions of the
form%
\begin{equation}
i\hslash \frac{\partial }{\partial s}\psi (s)=\frac{E}{c}\psi (s),
\label{stationarity approximation}
\end{equation}%
with $E$ being a real constant $4-$scalar independent of the proper time $s$%
. It follows that $\psi (s)$ is of the form%
\begin{equation}
\psi (s)=\exp \left\{ -\frac{i}{\hslash c}E(s-s_{o})\right\} \psi _{o}(g,%
\widehat{g},r),
\end{equation}%
being $\psi _{o}(g)\equiv \psi _{o}(g,\widehat{g},r)$ a\textbf{\ }solution
of the asymptotic proper time-independent quantum wave equation
\begin{equation}
\overline{H}_{R}^{(q)}\psi _{o}(g)=\frac{E}{c}\psi _{o}(g),
\label{time-independedt QWE}
\end{equation}%
to be referred to as \emph{stationary} \emph{vacuum CQG-wave equation. }%
Notice, in particular, that $\psi _{o}(g,\widehat{g},r)$\ here is assumed to
be suitably localized in the neighborhood of the background equilibrium
solution $\widehat{g}(r)\equiv \left\{ \widehat{g}_{\mu \nu }(r)\right\} $
so that possible additional classical stationary solutions can be
effectively ignored (see Part 1).

For definiteness, let us now invoke the perturbative approximation scheme
indicated above requiring in addition that the normalized effective quantum
potential density holding in the case of vacuum $\overline{V}\equiv
\overline{V}_{o}$ can be considered, at least in a suitable asymptotic
sense, as independent of $s$. It follows that in the subset of configuration
space set defined above, \textit{i.e}., the infinitesimal neighborhood $%
U_{g}(\widehat{g}(r),\varepsilon )$\ (see Eq.(\ref{neighborhood})) - upon
ignoring constant additive gauge contributions (to $\overline{H}_{R}^{(q)}$)
-\ Eq.(\ref{time-independedt QWE}) when expressed in terms of the field
variables $\delta g^{\mu \nu }$ takes the form:%
\begin{equation}
\frac{1}{2\alpha L}\left( -i\hslash \frac{\partial }{\partial \delta g_{\mu
\nu }}\right) \left( -i\hslash \frac{\partial }{\partial \delta g^{\mu \nu }}%
\right) \psi (s)-\frac{\sigma \alpha L\Lambda }{2}\left\{ \delta g_{\alpha
\beta }\delta g^{\alpha \beta }+\widehat{g}_{\alpha \beta }\delta g^{\alpha
\beta }\delta g^{\mu \nu }\widehat{g}_{\mu \nu }\right\} \psi (s)=0,
\label{qwe-particolare}
\end{equation}%
where by construction $-\frac{\sigma \alpha L\Lambda }{2}>0$. Next, let us
introduce the notations:%
\begin{equation}
\left\{
\begin{array}{c}
\delta g^{2}\equiv \delta g_{\mu \nu }\delta g^{\mu \nu }, \\
\delta \mathbf{\pi }^{2}\equiv \delta \pi _{\mu \nu }\delta \pi ^{\mu \nu },%
\end{array}%
\right.
\end{equation}%
with $\delta \pi ^{\mu \nu }$ identifying now the normalized quantum
canonical momentum operator $\delta \pi ^{\mu \nu }\equiv -\frac{i\hslash }{L%
}\frac{\partial }{\partial \delta g_{\mu \nu }}$ conjugate to the
displacement field $\delta g^{\mu \nu }$. Hence the same quantum wave
equation (\ref{qwe-particolare}) can be equivalently written in the form%
\begin{equation}
\left[ \frac{\delta \mathbf{\pi }^{2}}{2M}+\frac{1}{2}M\omega
^{2}L^{2}\left( \delta g^{2}+\widehat{g}_{\alpha \beta }\delta g^{\alpha
\beta }\delta g^{\mu \nu }\widehat{g}_{\mu \nu }\right) \right] \psi (s)=0,
\label{quantum-oscillator QWE}
\end{equation}%
to be referred to as \emph{quantum-oscillator quantum wave equation}, with
the operator%
\begin{equation}
H\equiv \frac{\delta \mathbf{\pi }^{2}}{2M}+\frac{1}{2}M\omega
^{2}L^{2}\left( \delta g^{2}+\widehat{g}_{\alpha \beta }\delta g^{\alpha
\beta }\delta g^{\mu \nu }\widehat{g}_{\mu \nu }\right)
\label{hamiltonian operator}
\end{equation}%
being referred to as (quantum)\emph{\ invariant-energy operator}. Moreover,
here $M$ and $\omega $ are the real $4-$scalars which respectively identify
the \emph{effective mass} and \emph{characteristic frequency }defined as%
\emph{\ }%
\begin{equation}
\left\{
\begin{array}{c}
M=\frac{\alpha }{cL}\equiv m_{o}, \\
\omega =c\sqrt{-\sigma \Lambda }.%
\end{array}%
\right.  \label{PHYSICAL PARAMETERS}
\end{equation}%
We conclude, therefore, that in the case of vacuum Eq.(\ref%
{quantum-oscillator QWE}) realizes in the same infinitesimal neighborhood $%
U_{g}(\widehat{g}(r),\varepsilon )$ the stationary CQG-wave equation
indicated above (see Eq.(\ref{time-independedt QWE})).

\section{6 - \emph{Application \#2 }- The discrete spectrum of the
stationary CQG-wave equation}

The eigenvalue equation (\ref{quantum-oscillator QWE}) is qualitatively
similar in form to the analogous quantum\ wave equation holding for the
quantum harmonic oscillator (QHO) in the case of ordinary quantum mechanics.
In this respect it must be clarified that the quadratic expansion of the
potential determined by Eq.(\ref{VACUUM POTENTIAL EXPANDED}) applies in the
neighborhood of the extremum set by the condition $\widehat{g}_{\mu \nu
}(r)=g_{\mu \nu }(r)$. The validity of Eq.(\ref{quantum-oscillator QWE}) is
physically supported by the fact that:

a)\ In the framework of the first-quantization approach adopted here,
quantum solutions are defined with respect to the classical background
represented by $\widehat{g}_{\mu \nu }(r)$\ only if the condition $\widehat{g%
}_{\mu \nu }(r)=g_{\mu \nu }(r)$\ is set to determine the extremum of the
potential. If this is not the case, one would have a harmonic expansion
around a solution which is not the stationary one, namely it is not the
classical solution of the Einstein equation. Other alternative extrema of
the potential remain necessarily excluded on such basis.

b) The condition $\widehat{g}_{\mu \nu }(r)=g_{\mu \nu }(r)$\ to set the
extremum of the harmonic expansion is the only physically acceptable one in
the conceptual framework developed here also because it yields the
stationary solution, thus motivating the search of eigenstates of the
stationary quantum harmonic oscillator given above.

c) The harmonic solution must be intended as a perturbative solution,
\textit{i.e}., it is an approximate local analytical solution of the
stationary quantum wave equation, and therefore it applies in a neighborhood
of the extremum point set by the condition $\widehat{g}_{\mu \nu }(r)=g_{\mu
\nu }(r)$.

The question which arises now is whether Eq.(\ref{quantum-oscillator QWE})
actually admits a discrete spectrum for its energy eigenvalues like QHO. In
this section we intend to prove that in validity of the prescription $%
-\sigma \Lambda >0$\ and based on Dirac's ladder operator approach, Eq.(\ref%
{quantum-oscillator QWE}) can be shown to admit indeed a discrete spectrum
of eigenvalues. For this purpose, let us introduce the creation and
annihilation operators for a spin-$2$\ particle, \textit{i.e}., represented
by second-order $4-$tensors$.$\ These can be identified respectively with
the operators $a_{\mu \nu }$ and $a_{\mu \nu }^{\dag }$:%
\begin{eqnarray}
a_{\mu \nu } &=&\sqrt{\frac{M\omega }{2\hslash }}\left( L\delta g_{\mu \nu }+%
\frac{L}{K_{o}}\widehat{g}_{\mu \nu }\delta g^{\alpha \beta }\widehat{g}%
_{\alpha \beta }-\frac{i}{M\omega }\delta \pi _{\mu \nu }\right) , \\
a_{\mu \nu }^{\dag } &=&\sqrt{\frac{M\omega }{2\hslash }}\left( L\delta
g_{\mu \nu }+\frac{L}{K_{o}}\widehat{g}_{\mu \nu }\delta g^{\alpha \beta }%
\widehat{g}_{\alpha \beta }+\frac{i}{M\omega }\delta \pi _{\mu \nu }\right) ,
\end{eqnarray}%
with $K_{o}$ denoting a suitable real number identified with one of the two
roots of the algebraic equation $K_{o}^{2}-2K_{o}-4=0$, namely $K_{o}=1\pm
\sqrt{5}$. Then one can show that in terms of the operator products $a_{\mu
\nu }a^{\dag \mu \nu }$ and $a_{\mu \nu }^{\dag }a^{\mu \nu }$ the identities%
\begin{equation}
H=\hslash \omega \left( a_{\mu \nu }a^{\dag \mu \nu }+\gamma \right) ,
\label{IDENTITY-1}
\end{equation}%
\begin{equation}
H=\hslash \omega \left( a_{\mu \nu }^{\dag }a^{\mu \nu }-\gamma \right) .
\label{IDENTITY-2}
\end{equation}%
hold, being $H$ the invariant-energy operator (\ref{hamiltonian operator})
and $\gamma $\textbf{\ }the constant real parameter $\gamma =8+\frac{2}{K_{o}%
}.$ The proof of both identities (\ref{IDENTITY-1}) and (\ref{IDENTITY-2})
follows from elementary algebra. Consider for example the proof of the first
one (\textit{i.e}., Eq.(\ref{IDENTITY-1})). The product $a_{\mu \nu }a^{\dag
\mu \nu }$ gives in fact%
\begin{equation}
a_{\mu \nu }a^{\dag \mu \nu }=\frac{M\omega }{2\hslash }\left( L^{2}\delta
g^{2}+L^{2}\left( \frac{4}{K_{o}^{2}}+\frac{2}{K_{o}}\right) \widehat{g}%
_{\alpha \beta }\delta g^{\alpha \beta }\delta g^{\mu \nu }\widehat{g}_{\mu
\nu }-\frac{\hslash }{M\omega }\left( 16+\frac{4}{K_{o}}\right) +\frac{1}{%
\left( M\omega \right) ^{2}}\delta \mathbf{\pi }^{2}\right) .
\end{equation}%
Hence, requiring that $\frac{4}{K_{o}^{2}}+\frac{2}{K_{o}}=1$ one obtains%
\begin{equation}
a_{\mu \nu }a^{\dag \mu \nu }=\frac{1}{2\hslash }\left( M\omega L^{2}\delta
g^{2}+M\omega L^{2}\widehat{g}_{\alpha \beta }\delta g^{\alpha \beta }\delta
g^{\mu \nu }\widehat{g}_{\mu \nu }+\frac{1}{M\omega }\delta \mathbf{\pi }%
^{2}\right) -\gamma ,
\end{equation}%
so that Eq.(\ref{IDENTITY-1}) manifestly applies. Furthermore, it is
immediate to obtain the following commutator relations:%
\begin{eqnarray}
\left[ a_{\mu \nu },a^{\dag \mu \nu }\right] &\equiv &-2\gamma , \\
\left[ a_{\mu \nu },a^{\dag \alpha \beta }\right] &\equiv &-\left( \delta
_{\mu }^{\alpha }\delta _{\nu }^{\beta }+\frac{1}{K_{o}}\widehat{g}^{\alpha
\beta }\widehat{g}_{\mu \nu }\right) ,
\end{eqnarray}%
so that the operator $H$ can equivalently be represented in the form
indicated by (\ref{IDENTITY-2}). Next, in terms of the operator $\mathcal{N}%
=a_{\mu \nu }^{\dag }a^{\mu \nu }$, elementary algebra shows that%
\begin{eqnarray}
\left[ \mathcal{N},a_{\mu \nu }\right] &=&a_{\mu \nu }+\frac{1}{K_{o}}%
\widehat{g}_{\mu \nu }\widehat{g}^{\alpha \beta }a_{\alpha \beta },
\label{COMMUTATOR-ID-1} \\
\left[ \mathcal{N},a_{\mu \nu }^{\dag }\right] &=&-a_{\mu \nu }^{\dag }-%
\frac{1}{K_{o}}\widehat{g}_{\mu \nu }\widehat{g}^{\alpha \beta }a_{\alpha
\beta }^{\dag }.  \label{COMMUTATOR-ID-2}
\end{eqnarray}%
In analogy with the Axioms given in Sections 2-3 and the adoption of a $4-$%
scalar wave function, it is possible to introduce also the $4-$scalar
operators $a=a_{\mu \nu }\widehat{g}^{\mu \nu }$\ and $a^{\dag }=a_{\mu \nu
}^{\dag }\widehat{g}^{\mu \nu }$ representing the projections of the tensor
operators $a_{\mu \nu }$ and $a_{\mu \nu }^{\dag }$ along the prescribed
metric tensor $\widehat{g}^{\mu \nu }$. Then, defining the normalized
operator $\widehat{\mathcal{N}}=N/\beta $,\ with $\beta \equiv 1+\frac{4}{%
K_{o}}$, it follows%
\begin{eqnarray}
\left[ \widehat{\mathcal{N}},a\right] &=&a, \\
\left[ \widehat{\mathcal{N}},a^{\dag }\right] &=&-a^{\dag },
\end{eqnarray}%
while identity (\ref{IDENTITY-2}) delivers for the invariant-energy operator
the representation\textbf{\ }%
\begin{equation}
H=\hslash \omega \left( \beta \widehat{\mathcal{N}}-\gamma \right) .
\label{INVARIANT-ENERGY OPERATOR}
\end{equation}

Therefore let us denote by $\left\vert \emph{n}\right\rangle $, $n$
respectively the eigenstate and real eigenvalue of the same operator $%
\widehat{\mathcal{N}}$, \textit{i.e}., such that $\widehat{\mathcal{N}}%
\left\vert n\right\rangle =n\left\vert n\right\rangle $, one notices that $%
\left\vert n\right\rangle $ is necessarily also an eigenvector of the
quantum energy operator $H$, with $n$ being generally not an integer number.
In addition, as a consequence of the commutator identities (\ref%
{COMMUTATOR-ID-1}) and (\ref{COMMUTATOR-ID-2}), it follows%
\begin{equation}
\widehat{\mathcal{N}}a\left\vert n\right\rangle =\left( a\widehat{\mathcal{N}%
}+\left[ \widehat{\mathcal{N}},a\right] \right) \left\vert n\right\rangle
=\left( a\widehat{\mathcal{N}}+a\right) \left\vert n\right\rangle =\left(
n+1\right) a\left\vert n\right\rangle ,
\end{equation}%
\begin{equation}
\widehat{\mathcal{N}}a^{\dag }\left\vert n\right\rangle =\left( a^{\dag }%
\widehat{\mathcal{N}}+\left[ \widehat{\mathcal{N}},a^{\dag }\right] \right)
\left\vert n\right\rangle =\left( a^{\dag }\widehat{\mathcal{N}}-a^{\dag
}\right) \left\vert n\right\rangle =\left( n-1\right) a^{\dag }\left\vert
n\right\rangle ,
\end{equation}%
which proves that the $4-$tensor operators $a_{\mu \nu }$\ and $a_{\mu \nu
}^{\dag }$\ act indeed as creation and annihilation operators. Therefore the
eigenvalues of the transformed states $a\left\vert n\right\rangle $ and $%
a^{\dag }\left\vert n\right\rangle $ are respectively $n+1$ and $n-1$, so
that introducing the corresponding eigenstates of the same operator, namely
such that
\begin{eqnarray}
\widehat{\mathcal{N}}\left\vert n+1\right\rangle &=&(n+1)\left\vert
n+1\right\rangle , \\
\widehat{\mathcal{N}}\left\vert n-1\right\rangle &=&(n-1)\left\vert
n-1\right\rangle ,
\end{eqnarray}%
one expects that $a^{\dag }\left\vert n\right\rangle =K_{(n+1)}\left\vert
n+1\right\rangle $ and $a\left\vert n\right\rangle =K_{(n-1)}\left\vert
n-1\right\rangle $, being $K_{(n+1)}$ and $K_{(n-1)}$ suitable $4-$scalars.
These results show that the spectrum of the\textbf{\ }non-negative
eigenvalues corresponding to the set of eigenstates%
\begin{equation}
\left\vert n-s\right\rangle ,\left\vert n-s+1\right\rangle ,..\left\vert
n\right\rangle ,\left\vert n+1\right\rangle ,\left\vert n+2\right\rangle ,..
\label{set of eigenstates}
\end{equation}%
is discrete and numerable, with the integer $k\equiv \min \left( s\right) $\
to be suitably prescribed. It is interesting to notice that in terms of the
operator $\widehat{\mathcal{N}}$\ defined above also the so-called number
operator can be prescribed, which by construction has only integer
eigenvalues. In fact, denoting by $n_{o}$ the minimum positive integer such
that $n_{o}\equiv int(n)\geq n$, then $N=\widehat{\mathcal{N}}\frac{n_{o}}{n}
$ identifies the so called\textbf{\ }\emph{number operator.} By construction
$N$\ has only integer eigenvalues, namely is such that $N\left\vert
n\right\rangle =n_{o}\left\vert n\right\rangle $, while being $N\left\vert
n-s\right\rangle =(n_{o}-s)\left\vert n-s\right\rangle $. Now we notice that
for all relative integers in the set $\left\{ -s,-s+1,...+\infty \right\} $\
the state $\left\vert n-s\right\rangle $\ is also an eigenstate of the
invariant-energy operator\ $H$.\ Its eigenvalue, referred to as
invariant-energy eigenvalue, is manifestly%
\begin{equation}
E_{n-s}\equiv \hslash \omega \left( \beta \left( n-s\right) -\gamma \right) .
\label{invariant quantum energy}
\end{equation}%
Then, the positive integer $k$\ in the set of eigenstates (\ref{set of
eigenstates}) can be prescribed in such a way that $E_{n-s}$\ has for $s=k$\
the minimum positive value, and hence identifies the \emph{minimum
invariant-energy eigenvalue}%
\begin{equation}
E_{\min }\equiv E_{n-k}=\hslash \omega \left( \beta \left( n-k\right)
-\gamma \right) .  \label{minimum invariant energy}
\end{equation}%
In terms of the integer $n_{o}$ indicated above one obtains for the minimum
positive eigenvalue of $H$,\ namely $E_{\min }\equiv E_{n-k}$ (see Eq.(\ref%
{minimum invariant energy})) the upper estimate%
\begin{equation}
E_{\min }<\hslash \omega \left( \beta \left( n_{o}-k\right) -\gamma \right)
\equiv \gamma _{o}\hslash \omega ,  \label{minimum  of the invariant energy}
\end{equation}%
where $\gamma _{o}$\ is the positive real number $\gamma _{o}=int(\gamma
)-\gamma >0$,\textbf{\ }and $int(\gamma )\equiv n_{o}-k$\ is the minimum
positive integer such that $\beta \left( n_{o}-k\right) >\gamma $.\ Since $%
K_{o}=1\pm \sqrt{5}$, then one can show that the only admissible admissible
root is the one associated with the positive-root, namely $K_{o1}\equiv
3.236 $. Hence, $\gamma =8+\frac{2}{1+\sqrt{5}}\equiv \gamma _{1}\cong 8.618$
and $\beta =1+\frac{4}{1+\sqrt{5}}\equiv \beta _{1}\cong 2.236$. This means
also that $\gamma _{o}\cong 0.326$, so that the majorization (\ref{minimum
of the invariant energy}) actually requires the \emph{weak upper bound }$%
E_{\min }\lesssim 0.326\hslash \omega $ to hold for the minimum
invariant-energy $E_{\min }$. In view of these considerations it follows
therefore that the stationary CQG-wave equation (\ref{quantum-oscillator QWE}%
) admits a spectrum of invariant-energy eigenvalues $E_{n}$ associated with
the quantum energy operator $H$ (see Eq.(\ref{hamiltonian operator})). The
minimum invariant energy $E_{\min }$ is non-vanishing as is proportional to
the characteristic frequency $\omega $ (see Eq.(\ref{minimum of the
invariant energy})). Finally, provided the cosmological constant $\Lambda $
is non-vanishing and $-\sigma \Lambda >0$, the spectrum indicated above is
discrete and numerable.

\section{7 - \emph{Application \#3} - Quantum prescription of the
characteristic scale length\emph{\ }$L(m_{o})$ and the graviton rest-mass $%
m_{o}$}

An interesting issue is related to the physical prescription at the quantum
level for the invariant parameter $L(m_{o})$ and consequently for the
rest-mass $m_{o}$ and the invariant parameter $\alpha $ appearing in the
quantum Hamiltonian operator $\overline{H}_{R}^{(q)}$ (see Eqs.(\ref{map-3}%
)). In principle these quantities are not necessarily the same as those
entering the corresponding classical normalized Hamiltonian structure $%
\left\{ \overline{x}_{R},\overline{H}_{R}\right\} $.

It should be stressed in fact that the precise prescription of the invariant
mass $m_{o}$ as well of $L(m_{o})$\ should follow from the theory of CQG
itself and be consistent with the physical interpretation of $m_{o}$\ in
terms of the graviton mass. A possible solution of the task can be achieved
based on the asymptotic treatment developed above in the case of vanishing
external fields and non-vanishing cosmological constant. It follows that, in
the perturbative framework considered here, both the mass prediction and the
invariant length therefore depend on actual (experimental or theoretical)
estimates for a possibly-non vanishing cosmological constant. We notice, in
fact, that the eigenvalue stationary quantum-wave equation (\ref%
{quantum-oscillator QWE}) and the related quantum energy operator (\ref%
{hamiltonian operator}) still depend both on the invariant mass $m_{o}$ and
the characteristic scale length $L\equiv L(m_{o})$. It must be remarked,
however, that the minimum energy prediction (\ref{minimum of the invariant
energy}) provides a possible prescription for the invariant mass $m_{o}$.
The minimum energy, or ground-state, eigenvalue $E_{\min }$, which is
associated with the quantum-oscillator quantum wave equation (\ref%
{quantum-oscillator QWE}), yields in fact the corresponding \emph{%
ground-state mass estimate }$m_{o}=\frac{E_{\min }}{c^{2}}$.

In view of Eq.(\ref{PHYSICAL PARAMETERS}) and the expression for $E_{\min }$
considered above, the upper bound estimate
\begin{equation}
m_{o}\lesssim 0.326\frac{\hslash \sqrt{-\sigma \Lambda }}{c}
\label{MASS-COSMOLOGICAL CONSTANT}
\end{equation}%
must apply, with the invariant rest-mass $m_{o}$ depending accordingly on
the cosmological constant. In case $-\sigma \Lambda >0$, the model
established by the stationary CQG-wave equation (\ref{quantum-oscillator QWE}%
) thus provides a tentative candidate for the identification of the
rest-mass of the massive graviton. Such a particle would therefore
necessarily be endowed with a sub-luminal speed of propagation. For the
example case considered above one can estimate the numerical value of $%
E_{\min }$ and $m_{o}$. Adopting for $\Lambda $ the current astrophysical
estimated value $\Lambda \cong 1.2\times 10^{-52}m^{-2}$ \cite%
{Planck-collaboration}, it is found that $E_{\min }\cong 1.1\times
10^{-52}J\sim 7\times 10^{-34}eV$, while $m_{o}\cong 1.26\times
10^{-69}kg\sim 7\times 10^{-34}eV/c^{2}$, so that the resulting
graviton-electron mass ratio is
\begin{equation}
\frac{m_{o}}{m_{e}}\cong 1.38\times 10^{-39},  \label{mass-ration}
\end{equation}%
with $m_{e}$ denoting the electron rest-mass.

The final problem to address pertains the identification of the invariant
length $L(m_{o})$ in the CQG-theory. If the graviton is considered as a
point-particle this can be identified either with the classical
Schwarzschild radius $L_{Sch}\equiv \frac{2Gm_{o}}{c^{2}}$\ associated with
the graviton rest mass $m_{o}$\ (see Part 1), or the Compton wavelength $%
\lambda _{C}\equiv \frac{\hbar }{m_{o}c}$. In the first case upon invoking
Eq.(\ref{MASS-COSMOLOGICAL CONSTANT}) one finds that $L_{Sch}\ll L_{Planck}$%
, where $L_{Planck}\cong 10^{-35}m$ is the Planck length. On the other hand
it is well known that the same Planck Length provides, at least in order of
magnitude, the minimum physically-admissible quantum length. Thus, "a
fortiori", it necessarily must realize a lower bound for the same
characteristic length $L(m_{o})$.\textbf{\ }This means that in the quantum
regime the classical prescription for $L(m_{o})$\ based on the Schwarzschild
radius is not physically acceptable, thus implying that quantum phenomena
for the graviton are dominant with respect to classical ones. Therefore the
prescription of $L(m_{o})$\ must be realized by means of the Compton
wavelength. In terms of Eq.(\ref{MASS-COSMOLOGICAL CONSTANT}) this yields in
the present case%
\begin{equation}
\lambda _{C}\equiv \frac{1}{\gamma _{o}}\frac{1}{\sqrt{-\sigma \Lambda }},
\label{compton-2}
\end{equation}%
while numerical evaluation of Eq.(\ref{compton-2}) gives $\lambda _{C}\cong
2.8\times 10^{26}m$. In view of the estimate for $E_{\min }$ this shows that
$L(m_{o})$\ necessarily coincides up to a factor of order unity with the
invariant characteristic length $L_{\Lambda }\equiv \frac{1}{\sqrt{-\sigma
\Lambda }}$ associated with the cosmological constant $\Lambda $, suggesting
at the same time also the possible quantum origin of the cosmological
constant \cite{Einstein1917,Winberg2000,Carroll2004}. The result is\ based
on the analytical estimate of minimum eigenvalue of the discrete spectrum
associated with the invariant-energy operator (the precise calculation of
the same eigenvalue can in principle be performed numerically). The
interpretation of $L_{\Lambda }$ in terms of graviton Compton wavelength
follows therefore on physical grounds and not simply on dimensional analysis
arguments.

In this framework, \emph{the cosmological constant }$\Lambda $\emph{\ is
associated with the Compton wavelength of the ground-state oscillation mode
of the quantum of the gravitational field, \textit{i.e}., the graviton mass }%
$m_{o}$\emph{. }As a final comment, it must be stressed that the estimate
given here for $m_{o}$ refers to the ground-state eigenvalue of the discrete
spectrum corresponding to the vacuum CQG-equation (\ref{quantum-oscillator
QWE}). However, each eigenvalue of the same discrete spectrum should give
rise to its corresponding mass value, so that the discrete energy spectrum
is sided by a discrete mass spectrum. On similar grounds, quantum wave
equations different from the vacuum one (\ref{quantum-oscillator QWE})
studied here should generate a corresponding different mass spectrum. Hence,
the value of $m_{o}$ is non-unique and depends both on the physical
properties of the background space-time as well as the solution spectrum of
the CQG wave equation to be solved (e.g., stationary or non-stationary
equation, vacuum or non-vacuum equation, etc.).

In connection with this we notice that, in the first-quantization approach
developed here, the metric tensor of the background space-time $\widehat{g}%
_{\mu \nu }(r)$, the Ricci curvature $4-$tensor $R_{\mu \nu }(\widehat{g}%
(r)) $ as well as the cosmological constant $\Lambda $ are all considered to
be in principle arbitrarily-prescribed quantities. The theory turns out to
be intrinsically background independent, \textit{i.e}., holding for any
realization of the space-time $\widehat{g}_{\mu \nu }(r)$. Nevertheless the
stationary approximation (\ref{stationarity approximation}) might not be
generally applicable, while even in the case in which the Hamiltonian
operator $\overline{H}_{R}^{(q)}$ does not depend explicitly on $s,$ the
proper time-dependent equation (\ref{QG-WAVW EQUATION}) may still admit
non-trivial explicitly time-dependent solutions. As a consequence a more
general class of solutions with respect to that considered above might occur
for the CQG-wave equation, as corresponds to complex physical
phenomenologies characterized by non-uniform behavior of both the prescribed
metric tensor $\widehat{g}_{\mu \nu }(r)$, of the corresponding Ricci tensor
$R_{\mu \nu }(\widehat{g}(r))$, as well as of the cosmological constant and
non-vanishing external fields.

As a final point, a peculiar connection exists between the classical
GR-Hamiltonian structure developed in Part 1 and the corresponding quantum
one represented by the quantum state $x^{(q)}\equiv \left( g_{\mu \nu
}^{(q)},\pi _{\mu \nu }^{(q)}\right) $ and the Hamiltonian operator $%
\overline{H}_{R}^{(q)}$. As pointed out in Part1, an arbitrary\ solution of
the GR-Hamilton equation, in particular the stationary solution $\widehat{g}%
_{\mu \nu }(r)$ determined by the vacuum Einstein field equations, is stable
with respect to infinitesimal perturbations provided suitable physical
conditions are met. As shown here, the same requirement applies in the
context of CQG-theory when the stationary quantum-wave equation Eq.(\ref%
{time-independedt QWE}) is reduced to a quantum harmonic oscillator. The
interesting consequence which emerges is therefore the validity of the
following mutual logical implication: the existence of a stable stationary
solution of the classical Hamiltonian structure of GR appears effectively,
at the same time, as a prerequisite and a consequence of the existence of a
discrete energy spectrum for the stationary CQG-wave equation, and hence
also the existence of a finite graviton rest-mass $m_{o}$.

\section{8 - Discussion and comparisons with literature}

Quantization methods, both in quantum mechanics and quantum gravity, are
usually classified in two approaches, the canonical and the covariant ones
\cite{hh1}. However, while for quantum mechanics the same approaches are
equivalent, this is not so in the case of quantum gravity. The reason is the
radically different approach taken in the two cases for the treatment of the
quantum state, the causality principle and of space-time itself.

In the canonical framework a canonical quantization approach is developed
which leaves formally arbitrary the space-time. Key ingredients usually
adopted for this purpose are, first, the introduction of ($3+1$)$-$ or ($2+2$%
)$-$decompositions \cite{zzz2,alcu,Vaca2,Vaca5,Vaca6}\textbf{\ }for the
representation of the same space-time and, second, the adoption of a quantum
state represented in terms of non-4-tensor continuum fields. In the
covariant approaches, instead, all physical quantities including the quantum
state are represented exclusively by means of $4-$tensor fields so that the
property of manifest covariance is automatically fulfilled. As a consequence
covariant quantization necessarily involves the assumption of some sort of
classical background space-time structure, for example identified with the
flat Minkowski space-time. In order to realize such a strategy, however, it
turns out that the quantum state is typically described by means of
superabundant variables. As a consequence covariant quantization usually
also requires the treatment of suitable constraint conditions.

Let us consider first the canonical approach. A choice of this type, for
example, is the one adopted by Dirac and based on the Dirac constrained
dynamics \cite{dirac4,dirac3,cast1,cast2,cast3}. By construction such an
approach is not manifestly covariant. We stress that this refers in
principle both to transformation properties with respect to local point
transformations, \textit{i.e}., LPT-theory, as well as the theory of
non-local point transformations (NLPT) developed in Ref.\cite{noi4}. It is
immediate to realize that this is indeed the case for the Dirac Hamiltonian
approach. In this picture in fact the field variable is identified with the
metric tensor $g_{\mu \nu }$, but the corresponding generalized velocity is
defined as $g_{\mu \nu ,0}$, namely with respect to the \textquotedblleft
time\textquotedblright\ component of the $4-$position. Consequently, in
Dirac's canonical theory, the canonical momentum remains identified with the
manifestly non-tensorial quantity $\pi _{Dirac}^{\mu \nu }=\frac{\partial
L_{EH}}{\partial g_{\mu \nu ,0}}$, where $L_{EH}$ is the Einstein-Hilbert
variational Lagrangian density.\ Hence, such a choice necessarily violates
the principle of manifest covariance \cite{noi1,noi2}.

The same kind of strategy was adopted in the approach developed later by
Arnowitt, Deser and Misner (ADM theory, 1959-1962 \cite{ADM}). In this case
manifest covariance is lost specifically because of the adoption, inherent
in the ADM approach, of Lagrangian and Hamiltonian variables which are not 4$%
-$tensors. In fact this is based on the introduction of the so-called 3+1
decomposition of space-time which, by construction is foliation dependent,
in the sense that it relies on a peculiar choice of a family of GR frames
for which time and space transform separately so that space-time is
effectively split into the direct product of a 1-dimensional time and a
3-dimensional space subsets respectively \cite{alcu}. For the same reason,
the quantum wave equation (\ref{QG-WAVW EQUATION}) proposed in this research
is intrinsically different from the Wheeler-DeWitt wave equation \cite{dew}.
In fact, Eq.(\ref{QG-WAVW EQUATION}) yields a dynamical evolution with
respect to the invariant proper-time $s$ defined on the background
space-time, while the Wheeler-DeWitt equation follows from the ADM foliation
theory and is expressed as an evolution Schr\"{o}dinger-like equation
advancing the dynamics of the wave function with respect to the
coordinate-time $t$, which is not an invariant parameter. In addition, in
the absence of background space-time, the same equation carries a conceptual
problem related to the definition of coordinate time, which is
simultaneously the dynamical parameter and a component of space-time which
must be quantized by solving the same equation. This problem however does
not arise in the theory of CQG proposed above.

Another interesting example worth to be mentioned is the one exemplified by
the so-called Ashtekar variables, originally identified respectively with a
suitable self-dual spinorial connection (the generalized coordinates) and
their conjugate momenta (see Refs.\cite{Ashtekar1986, Ashtekar1987}). It is
well-known that Ashtekar variables provide an alternative canonical
representation of SF-GR. Such a choice is at the basis of the so-called
\textquotedblleft loop representation of quantum general
relativity\textquotedblright\ \cite{Jacobson-Smolin1988} usually referred to
as "loop quantum gravity" (LQG) and first introduced by Rovelli and Smolin
in 1988-1990 \cite{Rovelli-Smolin1988,Rovelli-Smolin1990} (see also Ref.\cite%
{Rovelli1991}). Nevertheless, also the Ashtekar variables can be shown to be
by construction intrinsically non-tensorial in character either with respect
to the LPT or NLPT-groups. The basic consequence is that also the canonical
representation of Einstein field equations based on these variables, as well
as ultimately also LQG itself, violates the principle of manifest covariance.

Despite these considerations, it must be stressed that, as far as the
classical Hamiltonian formulation of GR is concerned, the canonical approach
and the manifestly-covariant theory proposed in Paper 1 and in the present
work are complementary, in that they exhibit distinctive physical properties
associated with two canonical Hamiltonian structures underlying GR itself.
The corresponding Hamiltonian flows, however, are different, being referred
to an appropriate coordinate-time of space-time foliation in the canonical
approach, and to a suitable invariant proper-time in the present theory. As
a consequence, the physical interpretation of quantum theories of GR build
upon these Hamiltonian structures remain distinctive. The CQG-theory
developed here in fact reveals the possible existence of a discrete spectrum
of metric tensors having non-vanishing momenta at quantum level, but whose
realization at classical level remains excluded for the extremal field
equations, \textit{i.e}., when the Hamiltonian theory is required to recover
the Einstein field equations (see also Paper 1 and further discussions on
the issue reported below in Section 9).

Let us now consider the covariant approaches to quantum gravity \cite%
{Ashtekar1974,Weinber1972,DeWitt1972}. In this case the usual strategy is to
split the space-time metric tensor $g_{\mu \nu }$ in two parts according to
the decomposition of the type $g_{\mu \nu }=\eta _{\mu \nu }+h_{\mu \nu }$,
where $\eta _{\mu \nu }$ is the background metric tensor defining the
space-time geometry (usually identified with the flat background), and $%
h_{\mu \nu }$ is the dynamical field (deviation field) for which
quantization applies. From the conceptual point of view there are some
similarities between the literature covariant approaches and the
manifestly-covariant quantum gravity theory developed here. The main points
of contact are: 1) the adoption of $4-$tensor variables, without invoking
any space-time foliation; 2) the implementation of a first-quantization
approach, in the sense that there exists by assumption a continuum classical
background space-time with a geometric connotation, over which the relevant
quantum fields are dynamically evolving; 3) the adoption of superabundant
variables, which in the two approaches are identified with the sets ($\eta
_{\mu \nu },h_{\mu \nu }$) and ($\widehat{g}_{\mu \nu },g_{\mu \nu }$)
respectively.

Nevertheless, important differences must be pointed out as well. In fact,
first of all, the CQG-theory developed here (and the CCG theory reported in
Part 1 on which it is based) is intrinsically non-perturbative in character.
It means, in other words, that the background metric tensor can be
identified with an arbitrary continuum solution of the Einstein equations,
while "\textit{a priori}" the canonical variable $g_{\mu \nu }$ is not
required to be a perturbation field. On the other hand, a decomposition of
the type (\ref{DECOMPOSITION FOR g}) resembling the one invoked in covariant
literature approaches can always be introduced "\textit{a posteriori}" for
the implementation of appropriate perturbative schemes. This occurs in
particular for analytical evaluation of discrete-spectrum quantum solutions
(see Sections 4-6 above). Second, the present theory is constructed starting
from the DeDonder-Weyl manifestly-covariant approach. As a consequence the
present approach is based on a variational formulation, which relies on the
introduction of a synchronous variational principle for the Einstein
equations first reported in Ref.\cite{noi1}. Such a feature is unique since
all previous literature is actually based on the adoption of asynchronous
variational principles, \textit{i.e.}, in which the invariant volume element
is considered variational rather than prescribed. Indeed, as shown in the
same reference it is precisely the synchronous principle which makes
possible the distinction between variational and extremal (or prescribed)
metric tensors, and the consequent introduction of non-vanishing canonical
momenta. As a result, manifestly-covariant classical Lagrangian, Hamiltonian
and Hamilton-Jacobi theories of GR have been formulated as pointed out in
Refs.\cite{noi1,noi2} and Part 1. In particular, based on the GR-Hamiltonian
structure determined in Part 1 such a feature allows for the adoption of the
canonical quantization represented by Eq.(\ref{canonical quantization}).
Third, in the present approach superabundant variables are implemented,
while the same covariant quantization holds with respect to a $4-$%
dimensional space-time, with no extra-dimensions being required for its
prescription. Fourth, the CQG-theory proposed in this paper is obtained from
the preliminary derivation of the reduced Hamiltonian theory given in Paper
1, which establishes a connection between field theory and particle
dynamics, although without requiring specification of any additional degrees
of freedom for the dynamical system predicted by Hamilton-Jacobi wave theory
\cite{FoP1,FoP2}. As a consequence the Hamiltonian structure on which
CCG-theory is built on is free of constraints, a feature which permits the
implementation of the canonical quantization\ rule represented by the same
equation (see again Eq.(\ref{canonical quantization})).

Finally, regarding covariant quantization, a further interesting comparison
concerns the Batalin-Vilkovisky (BV) formalism originally developed in Refs.%
\cite{BV2,BV3,BV4,BV5}. Such a method\ is usually implemented for the
quantization of gauge field theories and topological field theories in
Lagrangian formulation \cite{BV7,BV6,BV1}.\ Nevertheless also a
corresponding Hamiltonian formulation is available \cite{BFV1}. Further
critical aspects of the BV formalism can be found for example in Ref.\cite%
{BV8}. In the case of the gravitational field it has been formerly applied
in the context of perturbative quantum gravity to treat constraints arising
from initial metric decomposition (i.e., in reference with the so-called
gauge-fixing and ghost terms). Its basic features are the adoption of an
asynchronous Lagrangian variational principle of GR \cite{noi1},\ the use of
superabundant canonical variables and the consequent introduction of
constraints. These features mark the main differences with the present
manifestly-covariant Hamiltonian approach, which is non-perturbative,
follows from the synchronous Lagrangian variational principle defined in Ref.%
\cite{noi1} and is constraint-free. The key implication, as indicated in
Part 1, it that of permitting the construction of both the corresponding
Hamiltonian and Hamilton-Jacobi classical theories. These feature are
crucial also for the validity of the CQG-wave equation presented here (see
Eq.(\ref{QG-WAVW EQUATION})).

In conclusion, CQG-theory realizes at the same time a canonical and a
manifestly-covariant quantization method, in this way establishing a
connection both with\ former canonical and covariant approaches.
Nevertheless, the emerging new features of the present theory depart in
several ways from previous literature and might/should hopefully help
shading further insight into the long-standing problem of quantization of
gravity.

\section{9 - Conclusions}

The paper has been devoted to the formulation of a new theory of covariant
quantum gravity (CQG), referred to here as CQG-theory. The theoretical
foundations of the research presented here are based on the
manifestly-covariant Hamiltonian theory for the Einstein field equations
earlier developed in Refs.\cite{noi1,noi2,noi3}.

The quantum theory of gravitational field developed here distinguishes
itself from previous literature approaches to the problem. In fact, from one
side the present theory satisfies the principle of manifest covariance,
while at the same time the validity of the classical GR field equations is
preserved identically. Therefore, the realization of the CQG-theory does not
rely on the violation of manifest covariance in order to attempt a
quantization of the space-time through a discretization of its geometric
properties, nor it requires a modification of the Einstein field equations
at the variational level or the assumption "\textit{a priori}" of the
implementation of perturbative treatments from the start. The present theory
respects the canonical procedure well-known in the foundations of quantum
field theory, which requires to follow the logical path consisting in: a)
the identification of the appropriate classical Lagrangian density in $4-$%
scalar form; b) the subsequent definition of conjugate momenta and
realization of a corresponding classical Hamiltonian theory holding for a
canonical state; c) the introduction of canonical transformations and
development of Hamilton-Jacobi theory; d) the canonical quantization method
relying on classical Poisson brackets and the prescription of quantum wave
equation.

The development of the present CQG-theory is made possible by the adoption
of the new type of variational principle for the Einstein field equations,
for the first time pointed out in Ref.\cite{noi1}. The synchronous
variational formulation is characterized by distinguishing variational ($%
g_{\mu \nu }(r)$) and prescribed ($\widehat{g}_{\mu \nu }(r)$) tensor fields
in such a way that the variational ones are allowed to possess different
physical properties with respect to the prescribed fields, while preserving
at the same time the correct validity of the prescribed equations. In the
realm of the classical theory the physical behavior of variational fields
provide the mathematical background for the establishment of a
manifestly-covariant Hamiltonian theory of GR. The background metric tensor $%
\widehat{g}_{\mu \nu }(r)$ is purely classical and has a geometric
connotation, raising/lowering tensor indices and defining the Christoffel
symbols. At the classical level it must be $\widehat{\nabla }_{\alpha }%
\widehat{g}_{\mu \nu }(r)=0$, namely the covariant derivative of the
prescribed metric tensor is identically vanishing. Adopting the language of
classical dynamics, we can say by analogy that $\widehat{g}_{\mu \nu }(r)$
does not possess a "kinetic energy", since the corresponding generalized
"velocity field" $\widehat{\nabla }_{\alpha }\widehat{g}_{\mu \nu }(r)$ is
null by definition. However, the advantage of the synchronous variational
principle lies in the possibility of having variational metric tensor fields
$g_{\mu \nu }(r)$ for which the covariant derivative defined with respect to
the background space-time can be non-vanishing, so that $\widehat{\nabla }%
_{\alpha }g_{\mu \nu }(r)\neq 0$. We stress that this feature remains a
property of variational (and therefore virtual) fields $g_{\mu \nu }(r)$,
which therefore acquire a non-null generalized kinetic energy. This permits
the identification of canonical momenta and the construction of
corresponding covariant Hamiltonian theory holding for the Hamiltonian
structure $\left\{ x_{R},H_{R}\right\} $. When passing to the covariant
quantum theory variational fields become quantum observables and inherit the
corresponding tensor transformation laws of classical fields together with
the mentioned physical properties. It is then found that the quantum
observable corresponding to $g_{\mu \nu }(r)$ is endowed with non-vanishing
momenta having a quantum probability density. The resulting physical
interpretation of the present theory is straightforward. In the real of
classical theory the physical field $g_{\mu \nu }(r)$ is "frozen-in" with
the prescribed field $\widehat{g}_{\mu \nu }(r)$ which has a geometrical
connotation. Violation of the condition $\widehat{\nabla }_{\alpha }\widehat{%
g}_{\mu \nu }(r)=0$ is only allowed for variational fields. In the realm of
quantum theory the prescribed\textbf{\ }field $\widehat{g}_{\mu \nu }(r)$
keeps on retaining its meaning consistent with the picture of GR, while the
field $g_{\mu \nu }(r)$ acquires the physical meaning of a quantum field
which is permitted to deviate from $\widehat{g}_{\mu \nu }(r)$ and to
"oscillate" over the background space-time, thus violating at the quantum
level the frozen-in condition $\widehat{\nabla }_{\alpha }\widehat{g}_{\mu
\nu }(r)=0$. These features are exemplified by the structure of the
Hamiltonian density determined above, which can be expressed as the sum of
kinetic and potential density terms, in full analogy with standard quantum
theory of fields, as well as the possibility of recovering (at least in a
proper asymptotic treatment) the peculiar structure of the Hamiltonian
characteristic of the harmonic oscillator having a discrete spectrum of
eigenvalues.

The theory proposed here is believed to be susceptible of applications\ to a
wide range of quantum physics, theoretical physics and astrophysics-related
problems and to provide also new insight to\ the axiomatic foundations of
Quantum Gravity. Among the applications of CQG-theory special mention
deserve those which have been investigated in the paper. These include in
particular the proof of the existence of discrete energy spectra for the
stationary CQG-wave equation for solutions which are close to the classical
prescribed one $\widehat{g}_{\mu \nu }(r)$ and the related quantum
prescription of the free parameters which characterize the classical
Hamiltonian structure, namely the estimate for the graviton rest-mass $m_{o}$
and the determination of the characteristic scale length\ $L(m_{o})$.
Nevertheless, important issues concern also the search of more general
solutions pertaining to the non-stationary CQG-wave equation as well as
second-quantization effects such as the possible quantum modification of the
prescribed metric tensor associated with the background space-time. Such
tasks will be undertaken in forthcoming investigations.

\section{Acknowledgments}

Work developed within the research projects of the Czech Science Foundation
GA\v{C}R grant No. 14-07753P (C.C.) and the Albert Einstein Center for
Gravitation and Astrophysics, Czech Science Foundation No. 14-37086G (M.T.).

\end{document}